\documentclass[reprint,superscriptaddress,nofootinbib,aps,prd,10pt,preprintnumbers]{revtex4-1}

\usepackage{graphicx}
\usepackage{dcolumn}
\usepackage{bm}

\usepackage{amssymb,pslatex}
\usepackage{amsmath,marvosym}
\usepackage{braket}

\renewcommand{\d}{\ensuremath{\mathrm{d}}}
\newcommand{\p}{\partial}
\newcommand{\GZ}{\ensuremath{\mathrm{GZ}}}
\newcommand{\gf}{\ensuremath{\mathrm{gf}}}
\newcommand{\YM}{\ensuremath{\mathrm{YM}}}
\newcommand{\tr}{\mathop{\rm Tr}\nolimits}

\begin{document}

\title{{\Large {\bf Numerical Evaluation of the Bose-Ghost Propagator \\[2mm]
                    in Minimal Landau Gauge on the Lattice}}}

\author{Attilio~Cucchieri}
\email{attilio@ifsc.usp.br}
\affiliation{Instituto de F\'\i sica de S\~ao Carlos, Universidade de S\~ao
             Paulo, CP 369, 13560-970, S\~ao Carlos, SP, Brasil}
\author{Tereza~Mendes$^{1,}$}
\email{mendes@ifsc.usp.br}
\affiliation{Laboratoire de Physique Th\'eorique, CNRS, Univ.\ Paris-Sud 
et Universit\'e Paris-Saclay, B\^atiment 210, F-91405 Orsay Cedex, France}

\begin{abstract}
We present numerical details of the evaluation of the so-called Bose-ghost
propagator in lattice minimal Landau gauge, for the SU(2) case in four 
Euclidean dimensions. 
This quantity has been proposed as a carrier of the confining
force in the Gribov-Zwanziger approach and, as such, its infrared behavior
could be relevant for the understanding of color confinement in Yang-Mills
theories. Also, its nonzero value can be interpreted as direct evidence
of BRST-symmetry breaking, which is induced when restricting the functional
measure to the first Gribov region $\Omega$.
Our simulations are done for lattice volumes up to $120^4$ and for
physical lattice extents up to $13.5$ fm. 
We investigate the infinite-volume and continuum limits.

\preprint{LPT-Orsay-16-37}
\end{abstract}

\maketitle


\section{Introduction}

Restriction of the functional integral to the first Gribov region
$\Omega$ ---given as the set of transverse gauge configurations for which 
the Faddeev-Popov (FP) matrix ${\cal M}$ is non-negative--- defines the
so-called minimal Landau gauge in Yang-Mills theories \cite{Gribov:1977wm}.
On the lattice, this gauge fixing is implemented by a minimization
procedure (see e.g.\ \cite{Giusti:2001xf}),
without the need to consider the addition of a nonlocal horizon-function 
term $\gamma^4 S_h$ to the (Landau-gauge) action, as done in the 
Gribov-Zwanziger (GZ) approach in the continuum \cite{Zwanziger:1991ac}.
Let us recall that, in the GZ approach, the resulting (nonlocal) action may
be localized by introducing auxiliary fields. Also, one can define 
for these fields a
nilpotent BRST transformation (see e.g.\ \cite{Vandersickel:2012tz}),
which is a simple extension of the usual perturbative BRST transformation
that leaves the Yang-Mills action invariant in linear covariant gauge.
However, it can be easily verified that, in the GZ case, this local 
BRST symmetry
is broken by terms proportional to the Gribov parameter $\gamma$.
More precisely, since a nonzero value of $\gamma$ is related to the
restriction of the functional integration to $\Omega$, it is somewhat
natural to expect a breaking of the perturbative BRST symmetry, as a direct
consequence of the nonperturbative gauge-fixing. This issue has been
investigated in several works (see e.g.\ \cite{Zwanziger:2009je,
Zwanziger:2010iz,Maggiore:1993wq,Zwanziger:1993dh,vonSmekal:2008en,
Dudal:2008rm,Baulieu:2008fy,Dudal:2009xh,Sorella:2009vt,Kondo:2009qz,
Sorella:2010fs,Capri:2010hb,Dudal:2010hj,Capri:2011wp,Lavrov:2011wb,
Weber:2011nw,Lavrov:2012gb,Weber:2012vf,Maas:2012ct,Dudal:2012sb,Capri:2013naa,
Schaden:2013ffa,Reshetnyak:2013bga,Brambilla:2014jmp,Moshin:2014xka,
Capri:2014bsa,Schaden:2014bea,Reshetnyak:2014exa,Capri:2015ixa,Capri:2015nzw}
and references therein). The above interpretation is supported by the
recent introduction of a nilpotent nonperturbative BRST transformation
\cite{Capri:2015ixa,Capri:2015nzw}, which leaves the local
GZ action invariant.\footnote{To be more precise, before introducing the
nonperturbative BRST transformation, one also needs to rewrite the
horizon function in terms of a nonlocal gauge-invariant transverse
field $A_{\mu}^h(x)$ and redefine the Nakanishi-Lautrup field $b^a$
by a shift. See Refs.\ \cite{Capri:2015ixa,Capri:2015nzw} for
details.} The new symmetry is a simple modification of the usual
BRST transformation $s$, by adding (for some of the fields) a 
nonlocal term proportional to the Gribov parameter $\gamma$, thus 
recovering the usual perturbative transformation $s$ when $\gamma$ is 
set to zero.


As indicated above, the Gribov parameter $\gamma$ is {\em not}
introduced explicitly on the lattice, since in this case the restriction of 
gauge-configuration space to the region $\Omega$ is achieved directly by
numerical
minimization. Nevertheless, the breaking of the perturbative BRST symmetry
induced by the GZ action may be investigated by the lattice computation of
suitable observables, such as
the so-called Bose-ghost propagator, which has been proposed as a carrier
of the long-range confining force in minimal Landau gauge
\cite{Zwanziger:2009je,Furui:2009nj,Zwanziger:2010iz}.
The first numerical evaluation of the Bose-ghost propagator in minimal
Landau gauge was presented ---for the SU(2) case in four space-time 
dimensions--- in Refs.\ \cite{Cucchieri:2014via,Cucchieri:2014xfa}.
The data for this propagator show a strong infrared (IR) enhancement, with
a double-pole singularity at small momenta, in agreement with the one-loop
analysis carried out in Ref.\ \cite{Gracey:2009mj}. The data are also well
described by a simple fitting function, which can be related to a massive
gluon propagator in combination with an IR-free FP ghost
propagator. Those results constitute the first numerical manifestation of
BRST-symmetry breaking in the GZ approach.

Here we extend our previous calculations and present the details
of the numerical evaluation of the Bose-ghost propagator. In particular, 
we consider three different lattice definitions of this propagator
in order to check that the results obtained are in agreement with
each other. We also present our final data for the 
propagator ---complementing the results already reported 
in \cite{Cucchieri:2014via,Cucchieri:2014xfa}--- 
and we analyze the infinite-volume and continuum limits. 
The paper is
organized as follows. In the next section, following Refs.\
\cite{Cucchieri:2014via,Cucchieri:2014xfa,Vandersickel:2012tz}, we
set up the notation and introduce the continuum definitions necessary
for the evaluation of the Bose-ghost propagator. Then, in Section
\ref{sec:latt}, we address the corresponding definitions on the lattice and
describe the algorithms used in our Monte Carlo simulations. Finally, in the
last two sections, we present the output of these simulations and
our concluding remarks.


\section{The Bose-Ghost Propagator in the Continuum}

The localized GZ action\footnote{Here we follow the notation of the review
\cite{Vandersickel:2012tz}.} may be written for the general case of linear
covariant gauge as
\begin{equation}
S_{\GZ} \, = \, S_{\YM} + S_{\gf} + S_{\mathrm{aux}} + S_{\gamma} \; ,
\end{equation}
where the various terms correspond to
\begin{itemize}
\item the usual four-dimensional Yang-Mills action
\begin{equation}
S_{\YM} \, = \, \frac{1}{4} \, \int \d^{\rm 4} x \; F^a_{\mu \nu} \,
                                                    F^a_{\mu \nu} \; ;
\end{equation}
\item the covariant-gauge-fixing term
\begin{eqnarray}
\;\;\;\;\;\;\; S_{\gf} & = & \int \d^{\rm 4} x \; \Bigl[
              b^a \, \p_{\mu} A^a_{\mu} \,+\,
              \frac{\alpha}{2} \, b^a \, b^a \,+\,
              \overline{\eta}^{a} \, \p_{\mu} D^{ab}_{\mu} \, \eta^b                
              \Bigr] \; ,
\end{eqnarray}
which is parametrized by $\alpha$;
\item the auxiliary term
\begin{eqnarray}
\;\;\;\;\;\;\; S_{\mathrm{aux}} & = & \int \d^{\rm 4} x \,
\Bigl[ \overline{\phi}_{\mu}^{ac} \, \p_{\nu} \left( D_{\nu}^{ab}
                     \phi^{bc}_{\mu} \right) -
     \overline{\omega}_{\mu}^{ac} \, \p_{\nu} \left( D_{\nu}^{ab}
                     \omega^{bc}_{\mu} \right) \nonumber \\[1mm]
            &   & \;\;\; \qquad \qquad - g_0 \left( \p_{\nu}
         \overline{\omega}_{\mu}^{ac} \right) f^{abd} \,
                D_\nu^{be} \eta^e \phi_{\mu}^{dc} \Bigr] \; ,
\end{eqnarray}
which is necessary to localize the horizon function;
\item the $\gamma$ term
\begin{eqnarray}
\;\;\;\;\;\;\; S_{\gamma} & = & \int \d^{\rm 4}x \,
           \Bigl[ \gamma^{2}
                 D^{ab}_{\mu} \Bigl( \phi_{\mu}^{ab}
           + \overline{\phi}_{\mu}^{ab} \Bigr)
        - 4 \left( N_c^{2} - 1 \right) \gamma^{4} \Bigr] \; ,
\label{eq:Sofgamma}
\end{eqnarray}
which allows one to write the horizon condition as a stationary point
of the quantum action, yielding the gap equation.
\end{itemize}
In the above equations, 
$a,\,b,\, c,\,d$ and $e$ are color indices in the adjoint
representation of the SU($N_c$) gauge group, while $\mu$ and $\nu$ are
Lorentz indices. Repeated indices are always implicitly summed
over. Also,
\begin{equation}
F^a_{\mu \nu} \, = \, 
 \p_{\mu} \, A^a_{\nu} \, - \, \p_{\nu} \, A^a_{\mu} \, + \,
                       g_0 \, f^{abc} A^b_{\mu} A^c_{\nu}
\end{equation}
is the usual Yang-Mills field strength, we indicate with
\begin{equation}
D^{ab}_{\nu} \, = \, \delta^{ab} \p_{\nu}
                       + g_0 f^{acb} A_{\nu}^{c}\;,
\label{eq:Dcov}
\end{equation}
the covariant derivative in the adjoint representation, $A^b_{\mu}$ is
the gauge field, $g_0$ is the bare coupling constant and $f^{abc}$ are
the structure constants of the gauge group. At the same time, $b^a$ is
the Nakanishi-Lautrup field, (${\overline \eta}^b$, $\eta^b$) are the FP
ghost fields, $(\overline{\phi}^{ac}_{\mu}, \phi^{ac}_{\mu} )$ are
complex-conjugate bosonic fields and $(\overline{\omega}^{ac}_{\mu},
\omega^{ac}_{\mu} )$ are complex-conjugate Grassmann fields. 

In the limit $\alpha \to 0$, the above $S_{\GZ}$ action yields the usual
Landau gauge-fixing condition $\p_{\mu} A^a_{\mu} = 0$, while restricting
the functional integration to the region $\Omega$. For
a more detailed discussion of the GZ approach,
see e.g.\ \cite{Vandersickel:2012tz} and references therein.


\vskip 3mm

One defines the Bose-ghost propagator as \cite{Cucchieri:2014via,
Cucchieri:2014xfa}
\begin{equation}
Q^{abcd}_{\mu \nu}(x,y) \, = \,
\braket{ \,
 \omega^{ab}_{\mu}(x) \, \overline{\omega}^{cd}_{\nu}(y) \, + \,
 \phi^{ab}_{\mu}(x) \, \overline{\phi}^{cd}_{\nu}(y) \, } \; .
\label{eq:defBG}
\end{equation}
Let us note that this correlation 
function can also be written using the relation 
\begin{equation}
Q^{abcd}_{\mu \nu}(x,y) \, = \,
\braket{ \, s (\, \phi^{ab}_{\mu}(x) \,
\overline{\omega}^{cd}_{\nu}(y)) \, } \; .
\end{equation}
Here $s$ is the usual perturbative nilpotent BRST variation
\cite{Baulieu:1983tg}, which acts
on the fields entering the GZ action as
\begin{eqnarray}
\label{eq:sA}
s \, A^{a}_{\mu}                  &=& -\left( D_{\mu}\,\eta \right)^a
                        \; , \; \quad
s \, \eta^{a}                      =  \frac{1}{2} g_0 f^{abc}
                                      \eta^b \, \eta^c             \\[2mm]
s \, \overline{\eta}^{a}          &=& b^a                            
                        \; ,  \;\qquad \qquad \;
s \, b^{a}                         =  0                            \\[2mm]
s \, \phi^{ac}_{\mu}              &=& \omega^{ac}_{\mu}              
                        \; ,  \;\qquad \quad \;
s \, \omega^{ac}_{\mu}             =  0                            \\[2mm]
s \, \overline{\omega}^{ac}_{\mu} &=& \overline{\phi}^{ac}_{\mu}     
                        \; ,  \;\qquad \quad \; \; \,
s \, \overline{\phi}^{ac}_{\mu}    =  0 \; .
\label{eq:soverphi}
\end{eqnarray}
Under the above BRST transformations, one has
\cite{Vandersickel:2012tz}
\begin{equation}
s \, \left( S_{\YM} + S_{\gf} + S_{\mathrm{aux}} \right) \, = \, 0\;,
\end{equation}
while
\begin{equation}
s \, S_{\gamma} \propto \gamma^2 \neq 0 \; .
\label{eq:break}
\end{equation}
Thus, as mentioned in the Introduction,
the breaking of the perturbative BRST symmetry in the GZ approach is directly
related to a nonzero value of $\gamma$, i.e.\ it is induced by 
the restriction of gauge-configuration space to $\Omega$.

We see that the Bose-ghost propagator is written in terms of a BRST-exact 
quantity, which should have a
null expectation value for a BRST-invariant theory.
Note however that
it does not necessarily vanish if BRST symmetry is broken (see, for 
example, the discussion in Ref.\ \cite{Sorella:2009vt}). 
Let us recall that at tree
level (and in momentum space) one finds \cite{Vandersickel:2012tz,
Gracey:2010df}
\begin{equation}
\label{eq:Qabcd}
Q^{abcd}_{\mu \nu}(p,p') \, = \; \gamma^4 \,
 \frac{ \left(2 \pi\right)^4 \,\delta^{(4)}(p + p')
        \; g_0^2 \, f^{abe} f^{cde} \;P_{\mu \nu}(p)}{
          p^2 \, \left(p^4 + 2 g_0^2 \,N_c \gamma^4 \right)} \; ,
\end{equation}
where
\begin{equation}
P_{\mu \nu}(p) \, = \, \delta_{\mu \nu} \, - \,
          \frac{p_{\mu} p_{\nu}}{p^2}
\end{equation}
is the usual transverse projector. Thus, this propagator is proportional
to $\gamma$, i.e.\ its nonzero value is clearly
related to the breaking of the BRST symmetry in the GZ theory. As said
already in the Introduction, the above tree-level result has been extended
to one loop in Ref.\ \cite{Gracey:2009mj}.

Let us stress that our $Q^{abcd}_{\mu \nu}(x,y)$ propagator corresponds
to the $F$-term of the $V$-propagator of Ref.\ \cite{Zwanziger:2009je}
[see their Eqs.\ (72)--(75)]. One should also remark that the notations
used in Refs.\ \cite{Vandersickel:2012tz} and \cite{Zwanziger:2009je} are
slightly different. In particular, the factor $\gamma^4$ in the
former work is replaced by $\gamma$ in the latter one.


\vskip 3mm

On the lattice, one does not have direct access to the auxiliary fields
$(\overline{\phi}^{ac}_{\mu}, \phi^{ac}_{\mu} )$ and
$(\overline{\omega}^{ac}_{\mu}, \omega^{ac}_{\mu} )$. Nevertheless, since
these fields enter the continuum action at most quadratically, we can
integrate them out exactly, obtaining for the Bose-ghost propagator an
expression that is suitable for lattice simulations. More precisely, 
in order to arrive at such an expression for each of the two terms
in Eq.\ (\ref{eq:defBG}),
one can carry out the following steps \cite{Zwanziger:2009je}
\begin{enumerate}
\item add sources to the GZ action,
\item integrate out the auxiliary fields,
\item take the usual functional derivatives with respect to the sources,
      in order to obtain the chosen propagator.
\end{enumerate}
This procedure is, essentially, the inverse of the steps that allow
one to localize the horizon term. Indeed, in the localization process
one uses properties of Gaussian integrals (see, for example, Appendix A
in Ref.\ \cite{Vandersickel:2012tz}) for trading the nonlocal term in the
action\footnote{We refer to \cite{Vandersickel:2012tz} and references
therein for more details and subtleties in the definition of the
horizon function.}
\begin{eqnarray}
\!\!\!\! S_{\mathrm{nl}} & = & \gamma^4\,S_h \nonumber \\[2mm]
     &=&  \gamma^4 \,\int \d^{\rm 4} x \, \d^{\rm 4} y \;
                   D_{\mu}^{ac}(x) \, ( {\cal M}^{-1} )^{ab}(x,y) \,
                   D_{\mu}^{bc}(y)
\label{eq:Snl}
\end{eqnarray}
for the local one
\begin{eqnarray}
\!\!\!\!\!\!\!\!\!\! S_{\mathrm{l}} & = &
                - \int \d^{\rm 4} x \, \d^{\rm 4} y \;
      \overline{\phi}_{\mu}^{ac}(x) \, {\cal M}^{ab}(x,y) \,
                     \phi^{bc}_{\mu}(y) \nonumber \\[2mm]
     & & \; \; + \, \gamma^{2} \, \int \d^{\rm 4} x \;
                 D^{ac}_{\mu}(x) \Bigl[ \phi_{\mu}^{ac}(x)
           + \overline{\phi}_{\mu}^{ac}(x) \Bigr] \; ,
\label{eq:sl}
\end{eqnarray}
as can be easily checked by completing the quadratic form in the above
equation and by shifting the fields in the path integral. 
[Here, $D^{ac}_{\mu}(x)$ is the
covariant derivative, defined in Eq.\ (\ref{eq:Dcov}), and
${\cal M}^{ab}(x,y) = - \delta(x-y) \p_{\mu} D^{ac}_{\mu}(x)$ is the
FP matrix.] On the contrary, in the three-step procedure described above,
one starts from the local action $S_{\mathrm{l}}$ plus the source
terms
\begin{equation}
 \int \d^{\rm 4} x \;
                 J^{ab}_{\mu}(x) \, \phi_{\mu}^{ab}(x)
           + \overline{J}_{\mu}^{ab}(x)  \,
                   \overline{\phi}_{\mu}^{ab}(x)
\end{equation}
and ends up ---after integrating over $\overline{\phi}^{ab}_{\mu}$
and $\phi^{ab}_{\mu}$--- with the nonlocal expression
\begin{eqnarray}
& & \! \int \d^{\rm 4} x \, \d^{\rm 4} y \; \Bigl\{
       \Bigl[ \gamma^2 D_{\mu}^{ac}(x) + J^{ac}_{\mu}(x) \Bigr]
        \, ( {\cal M}^{-1} )^{ab}(x,y) \nonumber \\[2mm]
& & \qquad \qquad \qquad \; \; \Bigl[ \gamma^2 D_{\mu}^{bc}(y) +
                 \overline{J}_{\mu}^{bc}(y) \Bigr] \Bigr\} \; .
\end{eqnarray}
Then, by taking the functional derivative with respect to the sources
$J^{ab}_{\mu}(x)$ and $\overline{J}_{\nu}^{cd}(y)$, and by setting them
to zero, one obtains two terms for the propagator $ \braket{ \,
\phi^{ab}_{\mu}(x) \, \overline{\phi}^{cd}_{\nu}(y) \, }$. The first
term is simply $ \braket{ \, ( {\cal M}^{-1} )^{ac}(x,y) \, \delta^{bd}
\, \delta_{\mu \nu} \, }$ and it does not contribute to the Bose-ghost
propagator $Q^{abcd}_{\mu \nu}(x,y)$, since $ \braket{ \,
\omega^{ab}_{\mu}(x) \, \overline{\omega}^{cd}_{\nu}(y) \, }$ provides
an equal but opposite contribution.\footnote{This implies that the
behavior of the Bose-ghost propagator depends only on the bosonic fields
$(\overline{ \phi}^{ac}_{\mu}, \phi^{ac}_{\mu} )$.} The second term
yields
\cite{Zwanziger:2009je,Cucchieri:2014via,
Cucchieri:2014xfa}
\begin{equation}
\!\!\!\!\!\!\!\!
Q^{abcd}_{\mu \nu}(x-y) \, = \, \gamma^4 \, \left\langle \,
       R^{a b}_{\mu}(x) \, R^{c d}_{\nu}(y) \, \right\rangle \; ,
\label{eq:Qprop}
\end{equation}
where
\begin{equation}
R^{a c}_{\mu}(x) = \int \d^{\rm 4} z \,
         ( {\cal M}^{-1} )^{ae}(x,z) \, B^{ec}_{\mu}(z)
\label{eq:Rfunc}
\end{equation}
and $B^{ec}_{\mu}(z)$ is given by the covariant derivative
$D^{ec}_{\mu}(z)$. One can also note that, at the classical level, the
total derivatives $\p_{\mu} ( \phi_{\mu}^{aa} + \overline{ \phi}_{\mu}^{aa}
)$ in the action $S_{\gamma}$ 
---or, equivalently, in the second term of Eq.\ (\ref{eq:sl})--- can be
neglected \cite{Vandersickel:2012tz,Zwanziger:2009je}. 
In this case the expression for $B^{ec}_{\mu}(z)$ simplifies to
\begin{equation}
B^{ec}_{\mu}(z) \, = \, g_0 \, f^{e b c} \, A^{b}_{\mu}(z) \; ,
\label{eq:Bshort}
\end{equation}
as in Ref.\ \cite{Zwanziger:2009je}. Let us stress that, in both cases,
the expression for $Q^{abcd}_{\mu \nu}(x-y)$ in Eq.\ (\ref{eq:Qprop})
depends only on the gauge field $A^{b}_{\mu}(z)$ and can be evaluated
on the lattice. In fact, all auxiliary fields have been integrated out.

Finally, let us note that the above procedure is analogous to the lattice
evaluation of the ghost propagator
\begin{equation}
G^{ab}(x-y) \, = \, \left\langle \, \eta^{a}(x) \, \overline{\eta}^{b}(y)
                        \, \right\rangle \; .
\end{equation}
Indeed, also in this case, the Grassmann fields (${\overline \eta}^b$,
$\eta^b$) are not explicitly introduced on the lattice. Nevertheless, by
using the three-step procedure described above, one obtains the expression
\begin{equation}
G^{ab}(x-y) \, = \, \left\langle \,
( {\cal M}^{-1} )^{ab}(x,y) \, \right\rangle \; ,
\end{equation}
which can be considered in lattice numerical simulations 
(see for example Refs.\
\cite{Suman:1995zg,Cucchieri:1997dx}).


\section{Lattice Setup}
\label{sec:latt}

We evaluate the Bose-ghost propagator defined in Eqs.\
(\ref{eq:Qprop})--(\ref{eq:Rfunc}) above ---modulo the global factor
$\gamma^4$--- using Monte Carlo simulations applied to Yang-Mills
theory in four-dimensional Euclidean space-time
for the SU(2) gauge group.\footnote{As remarked in the Introduction,
the parameter $\gamma$ is not explicitly introduced on the lattice.
For this reason, quantities proportional to $\gamma$, such as the
Bose-ghost propagator considered here or the horizon function (see,
e.g., Ref.\ \cite{Cucchieri:1997ns}), are evaluated in lattice
simulations modulo the global $\gamma^4$ factor.}

In order to check for discretization effects, we considered four
different values of the lattice coupling $\beta$, namely \ 
$\beta_0 = 2.20$, 
$\beta_1 \approx 2.35$,
$\beta_2 \approx 2.44$
and $\beta_3 \approx 2.51$
respectively corresponding (see \cite{Fingberg:1992ju,Bloch:2003sk})
to a lattice spacing $a$ of about $0.21 \, fm$, $0.14\,
fm$, $0.11 \, fm$ and $0.08 \, fm$. These values are
summarized in Table \ref{tab:params}, while the lattice volumes $V$
considered for the various $\beta$'s are listed in Table \ref{tab:confs}.
Let us note that the sets of lattice volumes $V = 16^4$, $24^4$,
$32^4$, $40^4$, $48^4$ at $\beta_0$ and $V = 24^4$, $36^4$, $48^4$,
$60^4$, $72^4$ at $\beta_1$ yield (approximately) the same set of
physical volumes, ranging from about $(3.4 \, fm)^4$ to about $(10.1
\, fm)^4$. The lattice volumes $V = 96^4$ and $V = 120^4$, at
$\beta_2$ and $\beta_3$ respectively, also correspond to a physical
volume of about $(10.1 \, fm)^4$. Finally, the lattice volume
$V = 64^4$ at $\beta_0$ amounts to a physical volume of about
$(13.5 \, fm)^4$, which, at least in the study of the gluon
propagator, corresponds essentially to infinite volume
\cite{Cucchieri:2007md}. One should stress, however, that simulations up
to $V = 128^4$ at $\beta_0$ were necessary in order to achieve a clear
description of the IR behavior of the gluon propagator (see for example
Ref.\ \cite{Cucchieri:2011ig}). The lattice volumes $V = 64^4$ at
$\beta_0 $ and $V = 120^4$ at $\beta_3$ are new with respect to 
the data in Refs.\ \cite{Cucchieri:2014via,Cucchieri:2014xfa}.

Thermalized configurations have been generated using a standard heat-bath
algorithm accelerated by hybrid overrelaxation (see for example
\cite{Creutz:1980zw,Adler:1988gc}), with two overrelaxation sweeps for each
heat-bath sweep of the lattice. For the random number generator we use a
double-precision implementation of {\tt RANLUX} (version 3.2) with luxury
level set to two \cite{web}. The lattice minimal Landau gauge has been fixed
using the stochastic-overrelaxation algorithm \cite{Cucchieri:1995pn,
Cucchieri:1996jm,Cucchieri:2003fb} with a stopping criterion $( \p_{\mu}
\vec{A}_{\mu} )^2 \leq 10^{-14}$ (after averaging over the lattice volume
and over the three color components). 
As for the lattice gauge field $A_{\mu}(x)$,
corresponding to $a g_0 A_{\mu}(x)$ in the continuum, we employ the usual
unimproved definition $[U_{\mu}(x) - U^{\dagger}_{\mu}(x)]/(2 i)$, where
$U_{\mu}(x)$ are the lattice link variables entering the Wilson action.
Here, we did not check for possible Gribov-copy effects. All the relevant
parameters used for the numerical simulations can be found in Tables
\ref{tab:params} and \ref{tab:confs}.

In order to evaluate the Bose-ghost propagator, we invert the FP matrix
${\cal M}^{a b}(x,y)$ with the sources $B^{bc}_{\mu}(x)$, after removing
their zero modes. In our setup for the numerical simulations we follow
the notation described in Ref.\ \cite{Cucchieri:2005yr}. In particular,
for the action of the FP matrix on a color vector $v^b(x)$ we consider
Eq.\ (22) in \cite{Cucchieri:2005yr}, i.e.\
\begin{eqnarray}
\left( {\cal M} v \right)^b(x) & = & \sum_{\mu} \,
  \Gamma_{\mu}^{bc}(x) \left[ \, v^c(x) \,-\, v^c(x+e_{\mu}) \, \right]
                       \nonumber \\
  &  & \;\; + \, \Gamma_{\mu}^{bc}(x-e_{\mu}) \left[  \,v^c(x) \,-\,
            v^c(x-e_{\mu}) \, \right] \nonumber \\[2mm]
  &  & \;  \; \; + f^{b d c} \,
   \left[ \, A^d_{\mu}(x) \, v^c(x+e_{\mu}) \right.  \nonumber \\[2mm]
 & & \left. \;\;\; \; \; \; - \, A^d_{\mu}(x-e_{\mu}) \, v^c(x-e_{\mu})
                           \, \right] \; ,
\label{eq:Mdef}
\end{eqnarray}
where
\begin{equation}
\Gamma_{\mu}^{bc}(x) \,=\, \tr \, \left[  
       \left\{ \,\frac{\tau^b}{2}\,\mbox{,}\frac{\tau^c}{2}\,\right\} 
   \frac{U_{\mu}(x) + U^{\dagger}_{\mu}(x)}{2}
                 \right] \; .
\label{eq:defgamma}
\end{equation}
Here, we indicate with $\tau^b$ the $N_c^2 - 1$ traceless Hermitian generators
of the gauge group (in the fundamental representation) and with
$\left\{ , \right\}$ the usual anticommutator operation. In the SU(2)
case one has
\begin{equation}
\Gamma_{\mu}^{bc}(x) \, = \, \delta^{bc} \frac{\tr}{2} U_{\mu}(x)
\end{equation}
and $f^{b c d} \! = \! \epsilon^{b c d}$. At the same
time, for the sources, we allow for both possible definitions (see
discussion in the previous section), i.e.\ $B^{bc}_{\nu}(x)$ equal to the
covariant derivative $D^{bc}_{\nu}(x)$ and $B^{bc}_{\nu}(x) \, = \, g_0
\, f^{b e c} \, A^{e}_{\nu}(x)$. In the first case, by considering the close
relation between the nonlocal term contributing to the horizon function
[see Eq.\ (\ref{eq:Snl})] and the expression for the Bose-ghost propagator
[see Eqs.\ (\ref{eq:Qprop})--(\ref{eq:Rfunc})], it is natural to use as
sources the same functions $B^{bc}_{\mu}(x)$ entering the lattice
evaluation of the horizon function. Then, following Eq.\ (23) of Ref.\
\cite{Cucchieri:1997ns}, we have
\begin{eqnarray}
B^{bc}_{\mu}(x) &=& \left[ \Gamma_{\mu}^{bc}(x) \, - \,
                 \Gamma_{\mu}^{bc}(x-e_{\mu}) \right] \nonumber \\[2mm]
    &+& \! f^{c d b} \, \left[ \, A^d_{\mu}(x) \,+\, 
                                  A^d_{\mu}(x-e_{\mu}) \,\right] \; .
\label{eq:Blatt1}
\end{eqnarray}
In the second case, we consider two different discretizations of Eq.\
(\ref{eq:Bshort}), i.e.\
\begin{itemize}
\item the above Eq.\ (\ref{eq:Blatt1}) without the diagonal part in color
      space, i.e.\
\begin{equation}
B^{bc}_{\mu}(x) \, = \, f^{c d b} \, \left[ \, A^d_{\mu}(x) \,+\, 
                                  A^d_{\mu}(x-e_{\mu}) \,\right] \; ,
\label{eq:Blatt1nodia}
\end{equation}
\item and the trivial discretization
\begin{equation}
B^{bc}_{\mu}(x) \, = \, f^{b d c} \, A^d_{\mu}(x) \; .
\label{eq:Blatt2}
\end{equation}
\end{itemize}
Note that, in the continuum limit, there is a factor of -2 difference
between Eq.\ (\ref{eq:Blatt1nodia}) and Eq.\ (\ref{eq:Blatt2}), implying
a factor of 4 difference in the evaluation of the Bose-ghost propagator.


\begin{table}[t]
\protect\vskip 1mm
\begin{tabular}{| c | c | c |}
\hline
          & $\beta$ value       & $a$ (fm)
\\ \hline \hline
$\beta_0$ & $2.2$        & 0.21035
\\ \hline
$\beta_1$ & $2.34940204$ & 0.14023
\\ \hline
$\beta_2$ & $2.43668228$ & 0.10518
\\ \hline
$\beta_3$ & $2.50527693$ & 0.08414
\\ \hline
\end{tabular}
\caption{For each of our labelled choices of the lattice parameter $\beta$,
we give the value used in the simulations and the corresponding value of
the lattice spacing
$a$, in fm.
\label{tab:params}}
\end{table}


\begin{table}[b]
\protect\vskip 1mm
\begin{tabular}{| c | c | c | c | c | c | c | c |}
\hline
$\beta$    & $V$   & $L$ (fm)    & $\#$ confs & therm.\ & decorr.\ & $p$
& $\#$ CPUs
\\ \hline \hline
$\beta_0$ & $\,16^4$  & 3.366
& 10000       &  550       &  50      & 0.88 & 128
\\ \hline
$\beta_0$ & $\,24^4$  & 5.048
& 5000       &  770       &  70      & 0.91 & 256
\\ \hline
$\beta_0$ & $\,32^4$  & 6.731
& 1000       &  880       &  80      & 0.935 & 256
\\ \hline
$\beta_0$ & $\,40^4$  & 8.414
&  750       &  990       &  90      & 0.94 & 256
\\ \hline
$\beta_0$ & $\,48^4$  & 10.097
&  500       &  1100      &  110     & 0.95 & 256
\\ \hline
$\beta_0$ & $\,64^4$  & 13.462
& 300       &  1430      &  130     & 0.975 & 512
\\ \hline
$\beta_1$ & $\,24^4$  & 3.366
 & 5000       &  880       &  80      & 0.895  & 128 \\ \hline
$\beta_1$ & $\,36^4$  & 5.048
&  850       &  1100      &  100     & 0.915  & 216 \\ \hline
$\beta_1$ & $\,48^4$  & 6.731
&  500       &  1430      &  130     & 0.93   & 256\\ \hline
$\beta_1$ & $\,60^4$  & 8.414
&  400       &  1980      &  180     & 0.965  & 216 \\ \hline
$\beta_1$ & $\,72^4$  & 10.097
&  250       &  2000      &  200     & 0.975  & 256 \\ \hline
$\beta_2$ & $\,96^4$  & 10.097
&  100       &  2750      &  250     & 0.975  & 512 \\ \hline
$\beta_3$ & $\,120^4$ & 10.097
&  100       &  3000      &  300     & 0.975  & 500 \\ \hline
\end{tabular}
\caption{For each choice of the coupling $\beta$ (see values in Table
\ref{tab:params}) and lattice volume $V$,
we indicate the lattice extent in physical units $L$, 
the number of configurations considered, the number of
thermalization sweeps used to generate the first configuration
(starting from a random initial configuration), the number of
decorrelation sweeps (between two thermalized configurations), the
value of the parameter $p$ used in the stochastic overrelaxation
algorithm, and the number of (quadcore) {\tt Blue Gene/P} CPUs
used for the simulations.
\label{tab:confs}}
\protect\vskip -0.7cm
\end{table}

\begin{figure}[t]
\centering
\vskip -2mm
\includegraphics[trim=55 0 40 0, clip, scale=1.05]{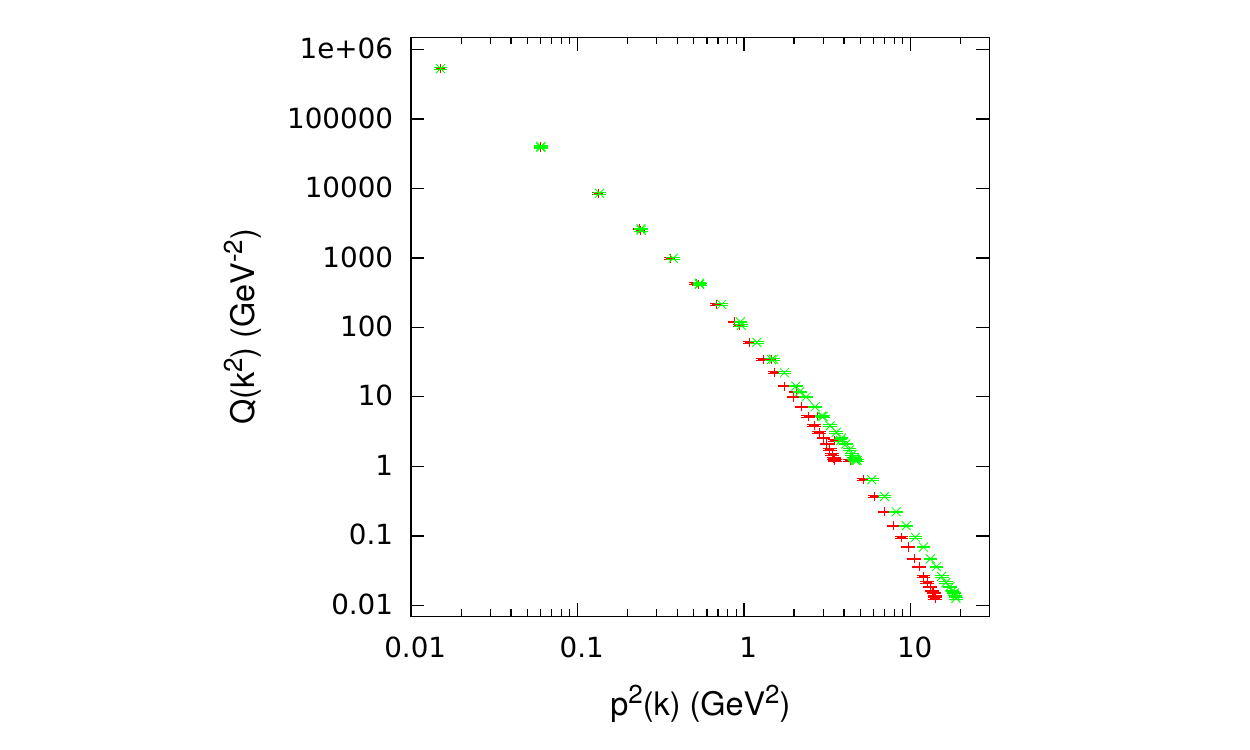}
\caption{\label{fig:rot0} The Bose-ghost propagator $Q(k^2)$,
defined in Eq.\ (\ref{eq:Q}), as a function of the unimproved
(red, $+$) and of the improved (green, $\times$) momentum squared
$p^2(k)$ [see Eqs.\ (\ref{eq:pun})--(\ref{eq:pk})]. We plot data for
$\beta_0$ and $V = 48^4$ using the sources defined in Eq.\
(\ref{eq:Blatt1}). Note the logarithmic scale on both axes.
}
\end{figure}

\begin{figure}[t]
\centering
\vskip -2mm
\includegraphics[trim=55 0 40 0, clip, scale=1.05]{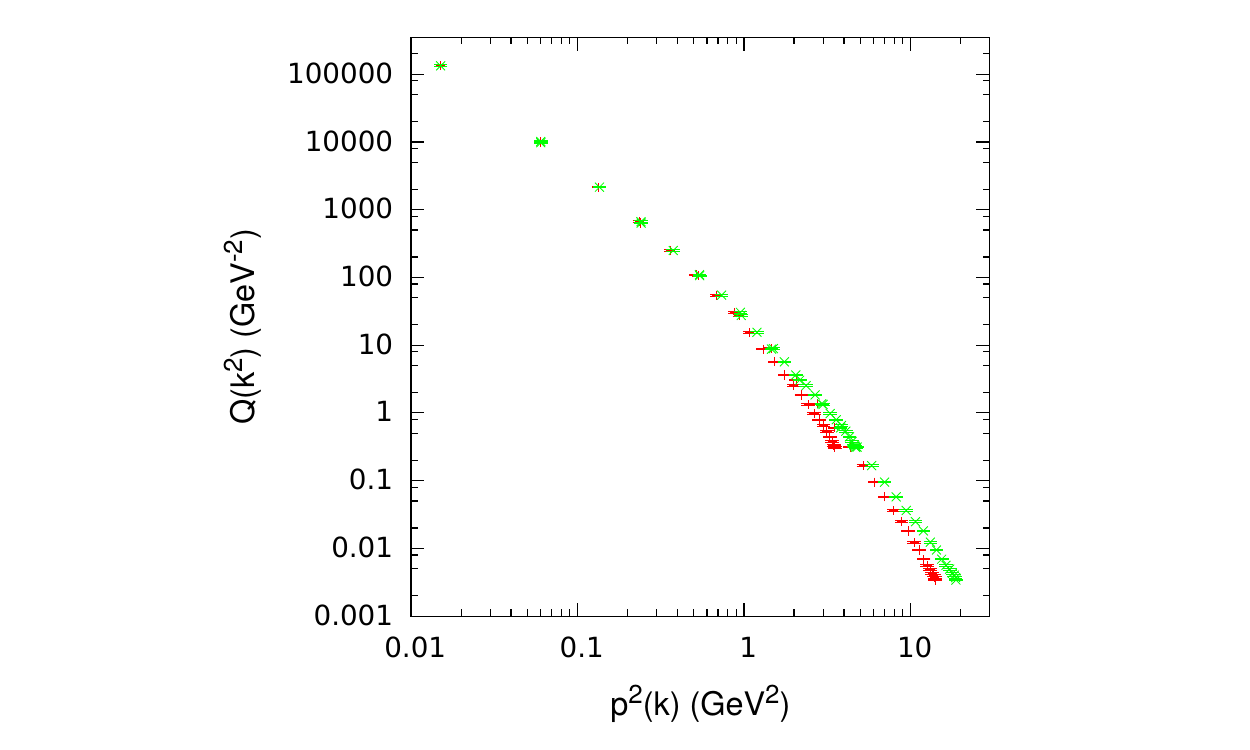}
\caption{\label{fig:rot} The Bose-ghost propagator $Q(k^2)$,
defined in Eq.\ (\ref{eq:Q}), as a function of the unimproved
(red, $+$) and of the improved (green, $\times$) momentum squared
$p^2(k)$ [see Eqs.\ (\ref{eq:pun})--(\ref{eq:pk})]. We plot data for
$\beta_0$ and $V = 48^4$ using the sources defined in Eq.\
(\ref{eq:Blatt2}). Note the logarithmic scale on both axes.
}
\end{figure}

\begin{figure}[t]
\centering
\vskip -2mm
\includegraphics[trim=55 0 40 0, clip, scale=1.05]{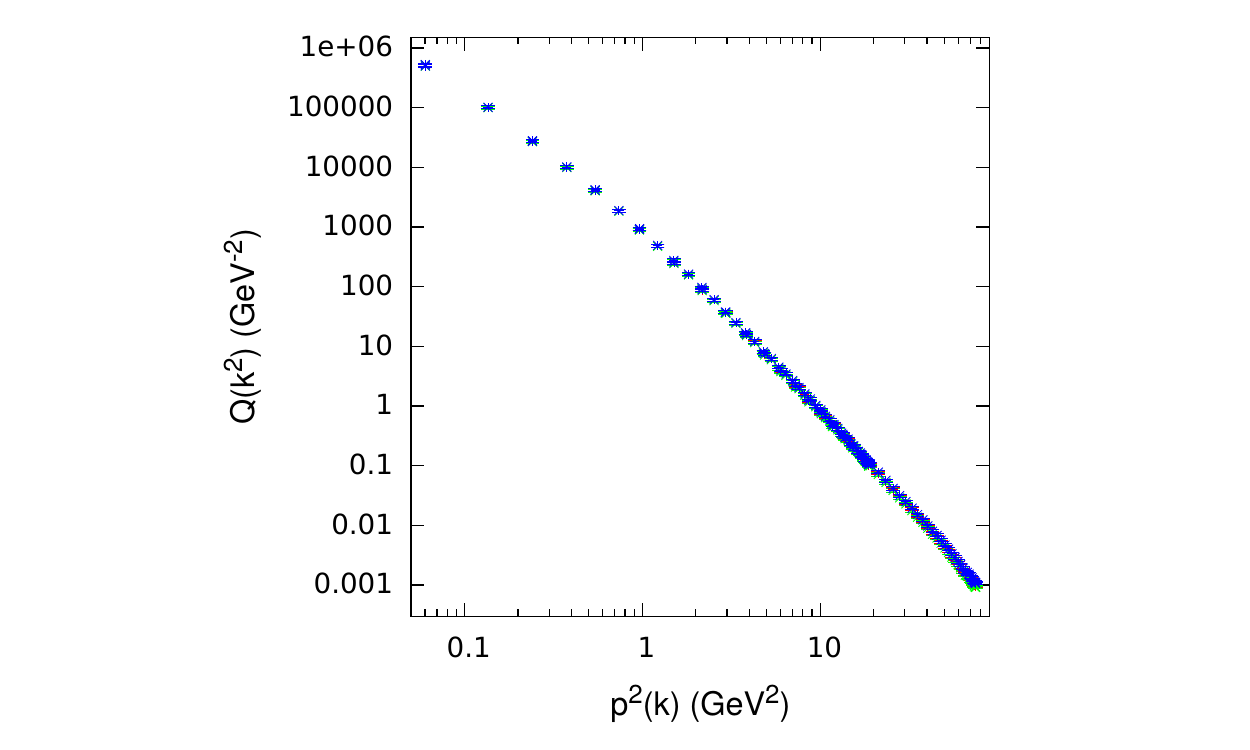}
\caption{\label{fig:dis} The Bose-ghost propagator $Q(k^2)$,
defined in Eq.\ (\ref{eq:Q}), as a function of the improved
momentum squared $p^2(k)$ [see Eqs.\ (\ref{eq:pim}) and (\ref{eq:pk})]
for the lattice volume $V = 96^4$ at $\beta_2$. Here we
plot data for the first (red, $+$), second (green, $\times$) and
third (blue, $*$) proposed discretizations of the sources
$B^{bc}_{\mu}(x)$ [see Eqs.\ (\ref{eq:Blatt1})--(\ref{eq:Blatt2}) in
Section \ref{sec:latt}]. For the last case the data are multiplied
by a factor 4. Note the logarithmic scale on both axes.
}
\end{figure}

\begin{figure}[t]
\centering
\vskip -2mm
\includegraphics[trim=55 0 40 0, clip, scale=1.05]{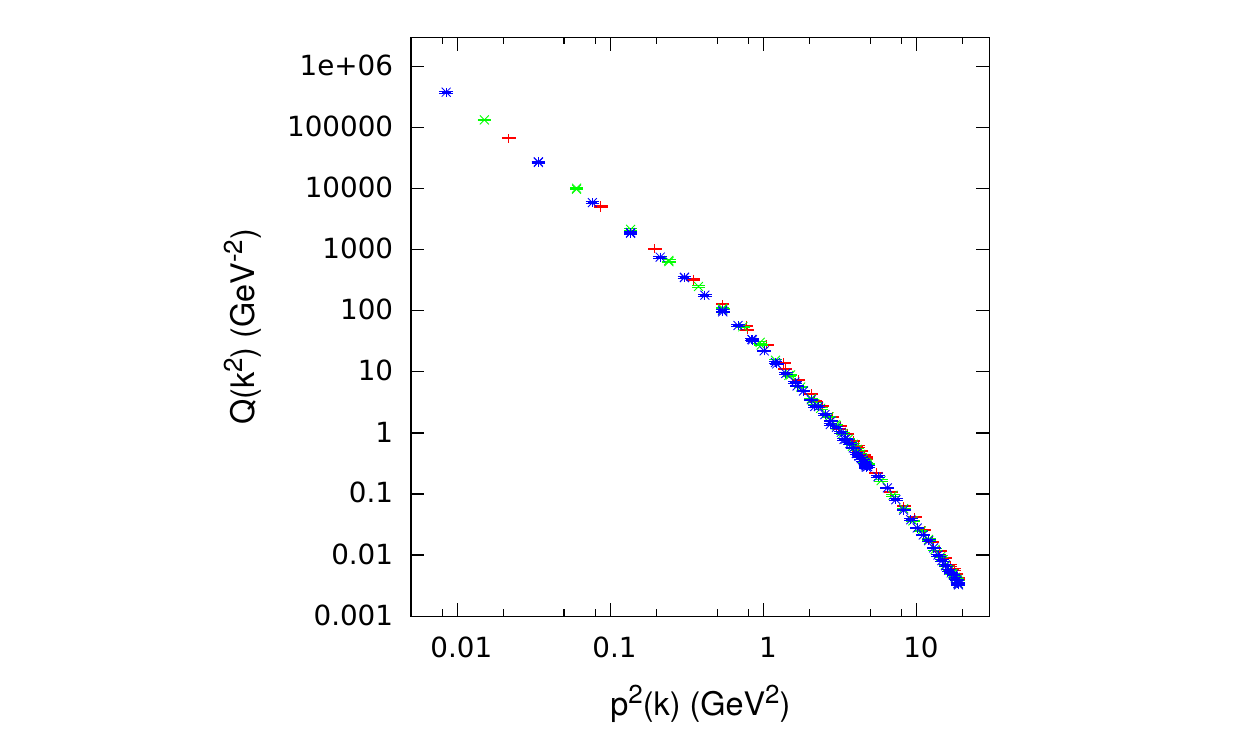}
\caption{\label{fig:vol2} The Bose-ghost propagator $Q(k^2)$,
defined in Eq.\ (\ref{eq:Q}), as a function of the improved
momentum squared $p^2(k)$ [see Eqs.\ (\ref{eq:pim}) and (\ref{eq:pk})].
We plot data for $\beta_0$ and $V = 40^4$ (red, $+$), $48^4$
(green, $\times$), $64^4$ (blue, $*$), using the sources defined in
Eq.\ (\ref{eq:Blatt2}). Note the logarithmic scale on both axes.
}
\end{figure}

\begin{figure}[t]
\centering
\vskip -2mm
\includegraphics[trim=55 0 40 0, clip, scale=1.05]{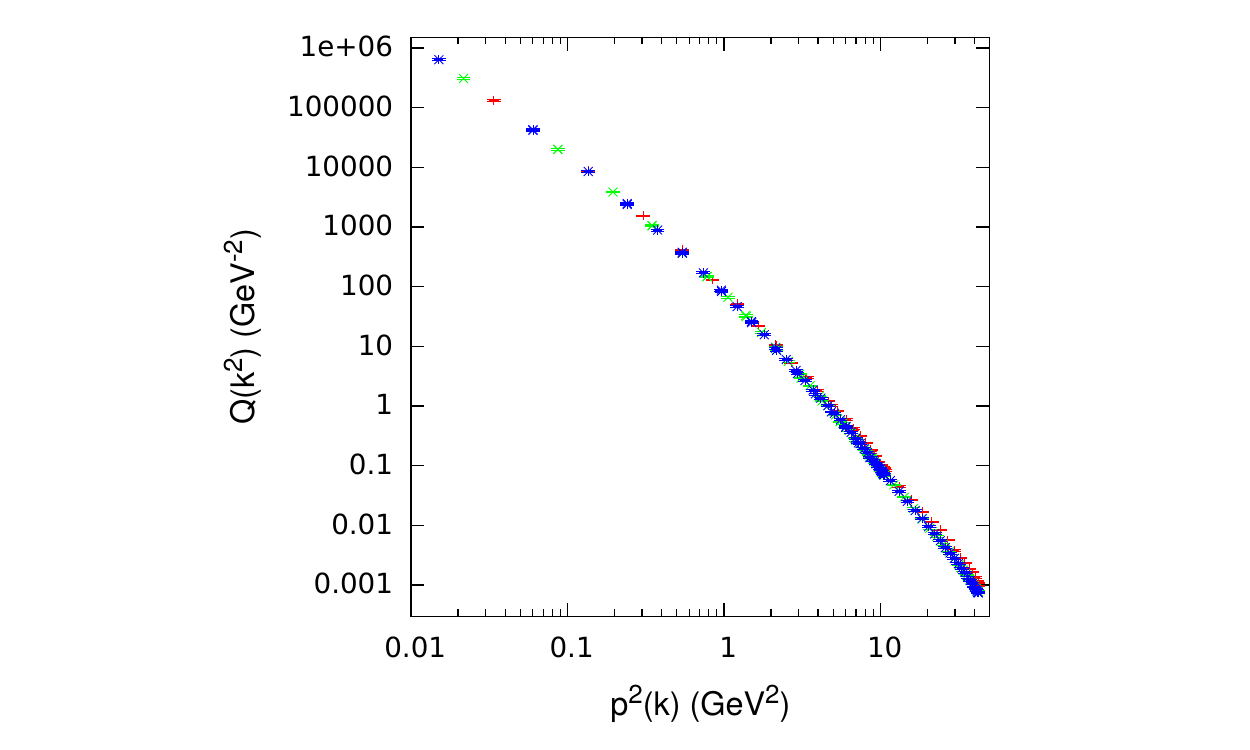}
\caption{\label{fig:vol} The Bose-ghost propagator $Q(k^2)$,
defined in Eq.\ (\ref{eq:Q}), as a function of the improved
momentum squared $p^2(k)$ [see Eqs.\ (\ref{eq:pim}) and (\ref{eq:pk})].
We plot data for $\beta_1$ and $V = 48^4$ (red, $+$), $60^4$
(green, $\times$), $72^4$ (blue, $*$), using the sources defined in
Eq.\ (\ref{eq:Blatt2}). Note the logarithmic scale on both axes.
}
\end{figure}

\begin{figure}[t]
\centering
\vskip -2mm
\includegraphics[trim=55 0 40 0, clip, scale=1.05]{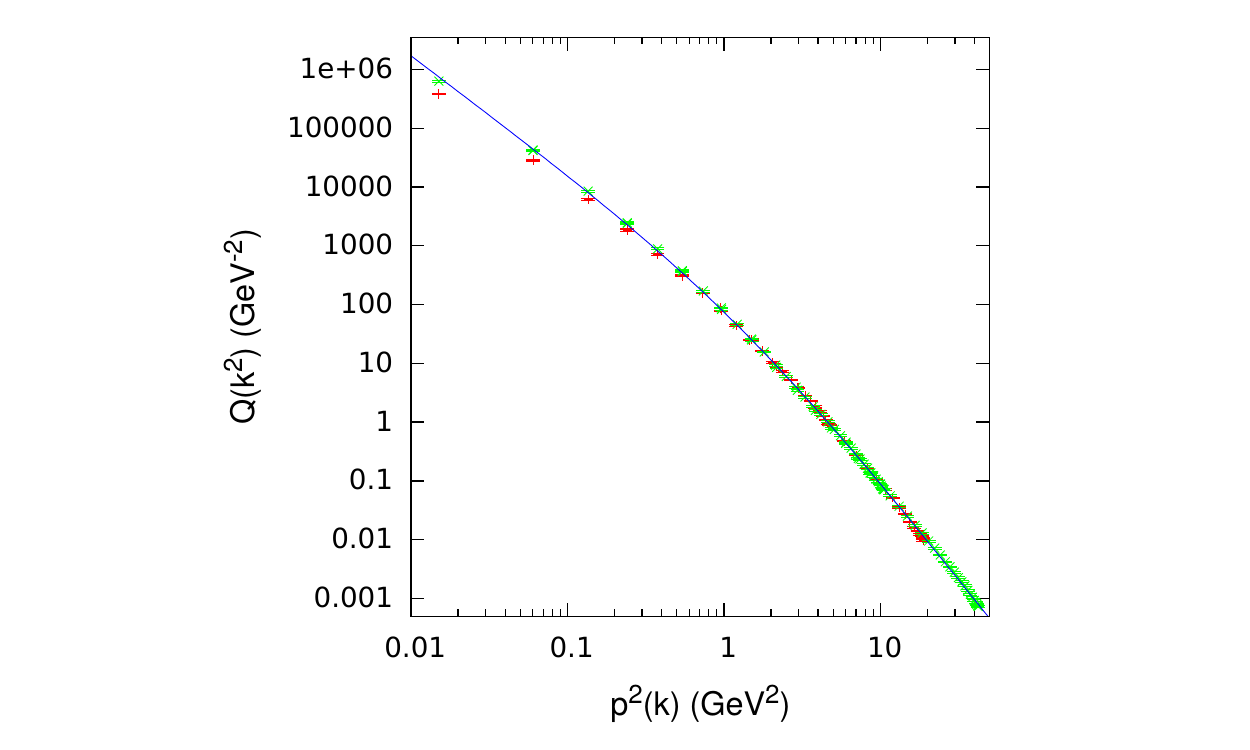}
\vskip 1mm
\includegraphics[trim=55 0 40 0, clip, scale=1.05]{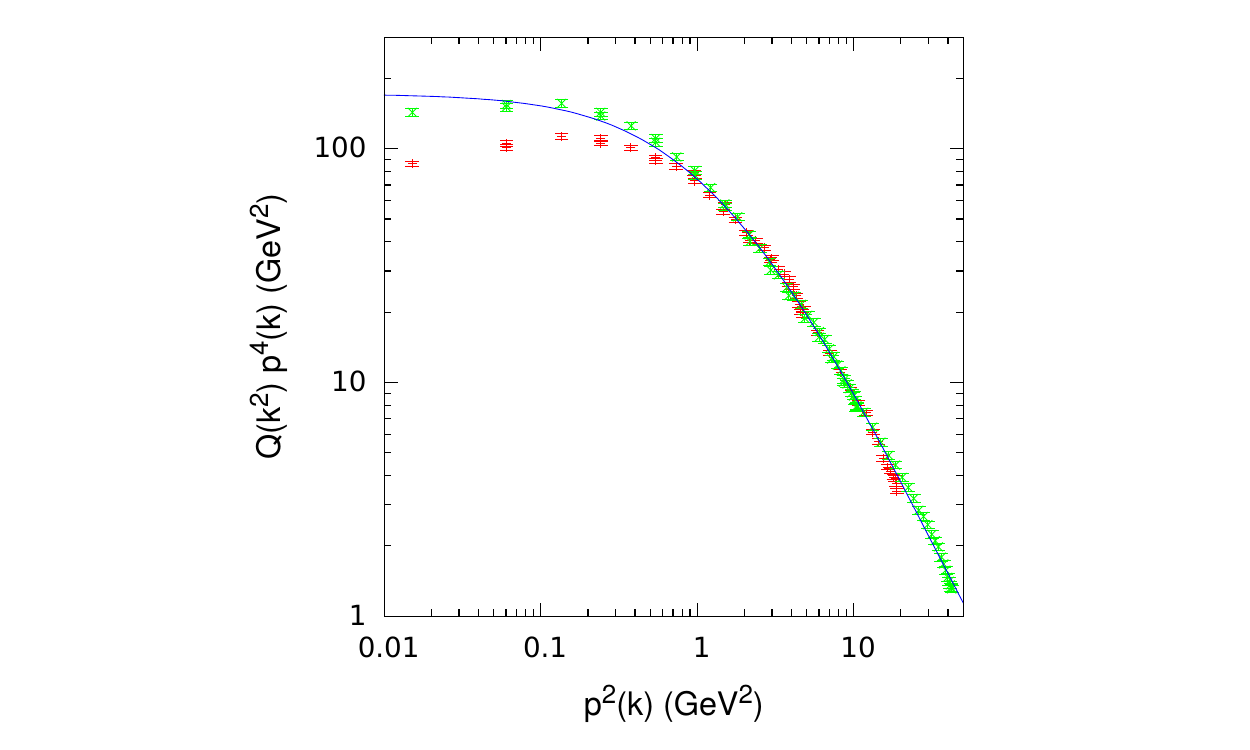}
\caption{\label{fig:fit} The Bose-ghost propagator $Q(k^2)$ ({\bf top})
[see Eq.\ (\ref{eq:Q})] and the rescaled Bose-ghost propagator
$Q(k^2) p^4(k)$ ({\bf bottom}), as a function of the improved momentum
squared $p^2(k)$ [see Eqs.\ (\ref{eq:pim}) and (\ref{eq:pk})].
We plot data for $\beta_0$, $V = 48^4$ (red, $+$) and $\beta_1$,
$V = 72^4$ (green, $\times$), after applying a matching
procedure \cite{Leinweber:1998uu,Cucchieri:2003di} to the former set of
data. Here we use the sources defined in Eq.\ (\ref{eq:Blatt2}).
We also plot, for $V = 72^4$, a fit using Eq.\ (\ref{eq:fit}) and the
parameters in Table \ref{tab:fits}, with $c = 37(4)$. Note the logarithmic
scale on both axes.
}
\end{figure}

\begin{figure}[t]
\centering
\vskip -2mm
\includegraphics[trim=55 0 40 0, clip, scale=1.05]{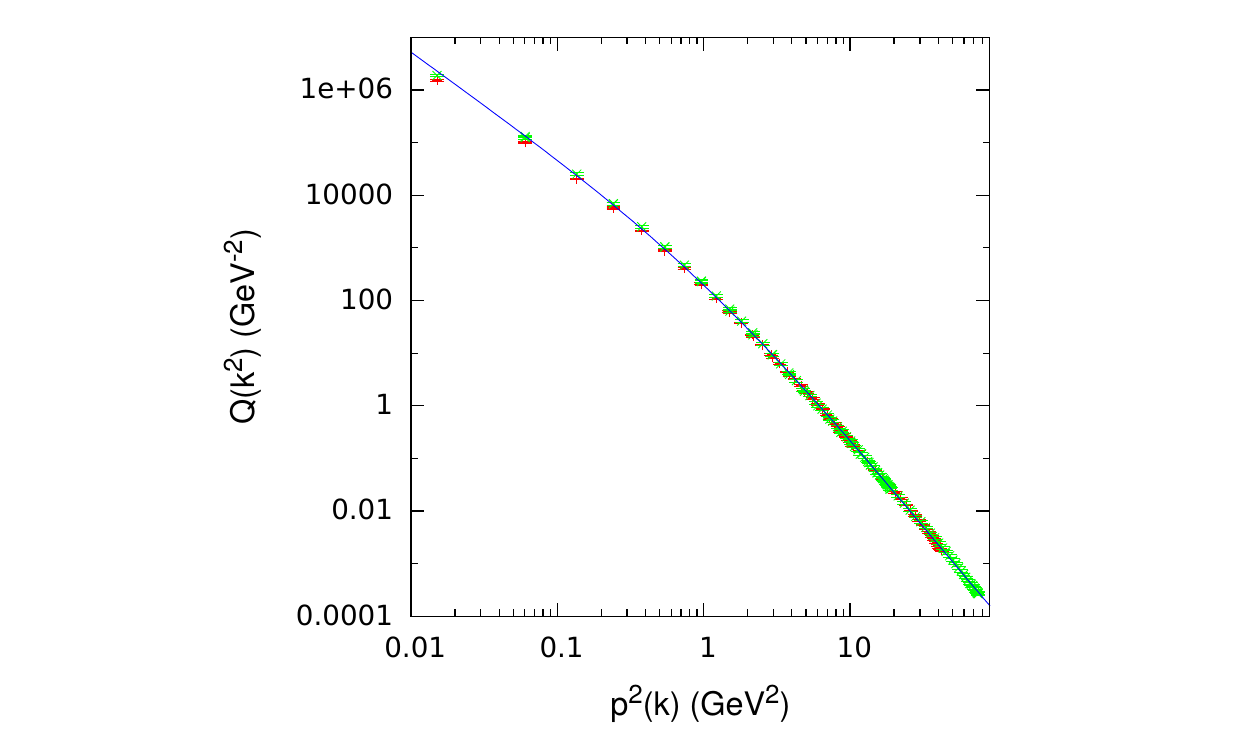}
\vskip 1mm
\includegraphics[trim=55 0 40 0, clip, scale=1.05]{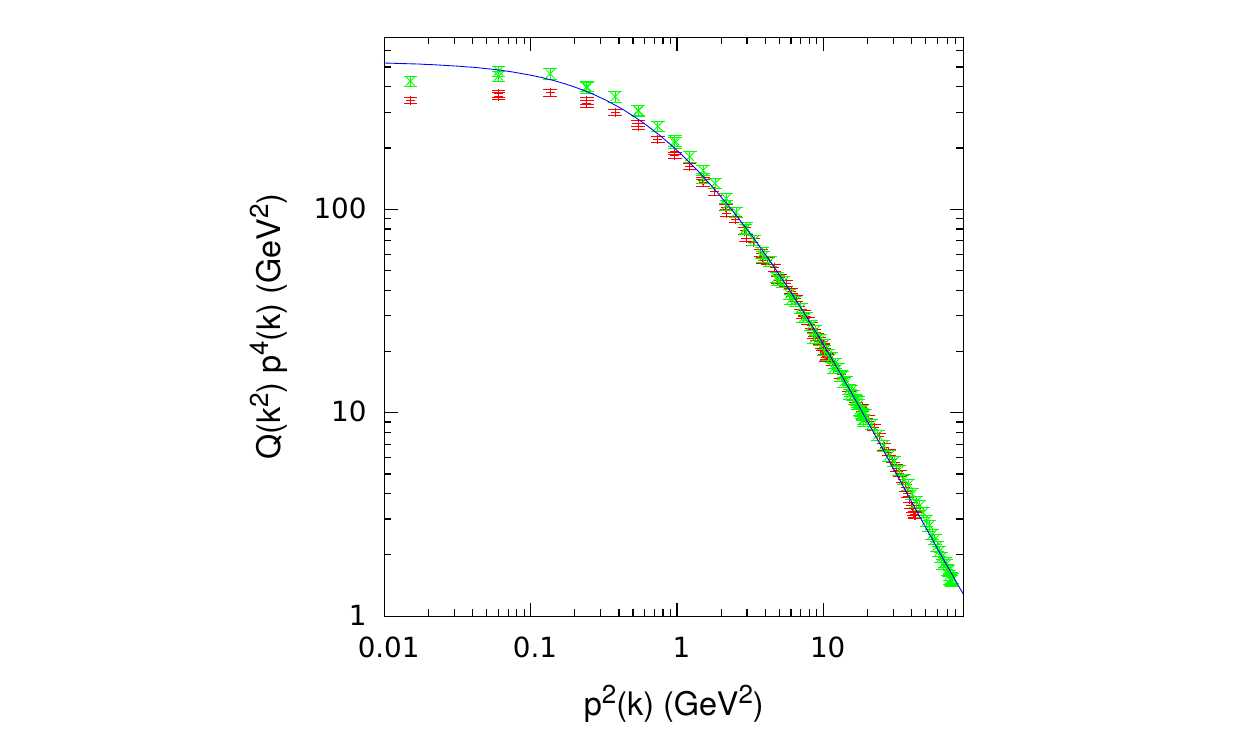}
\caption{\label{fig:fit2} The Bose-ghost propagator $Q(k^2)$ ({\bf top})
[see Eq.\ (\ref{eq:Q})] and the rescaled Bose-ghost propagator
$Q(k^2) p^4(k)$ ({\bf bottom}), as a function of the improved momentum
squared $p^2(k)$ [see Eqs.\ (\ref{eq:pim}) and (\ref{eq:pk})].
We plot data for $\beta_1$, $V = 72^4$ (red, $\times$) and
$\beta_2$, $V = 96^4$ (green, $*$), after applying a matching
procedure \cite{Leinweber:1998uu,Cucchieri:2003di} to the former set of
data. Here we use the sources defined in Eq.\ (\ref{eq:Blatt2}).
We also plot, for $V = 96^4$, a fit using Eq.\ (\ref{eq:fit}) and the
parameters in Table \ref{tab:fits}, with $c = 82(5)$. Note the logarithmic
scale on both axes.
}
\end{figure}

\begin{figure}[t]
\centering
\vskip -2mm
\includegraphics[trim=55 0 40 0, clip, scale=1.05]{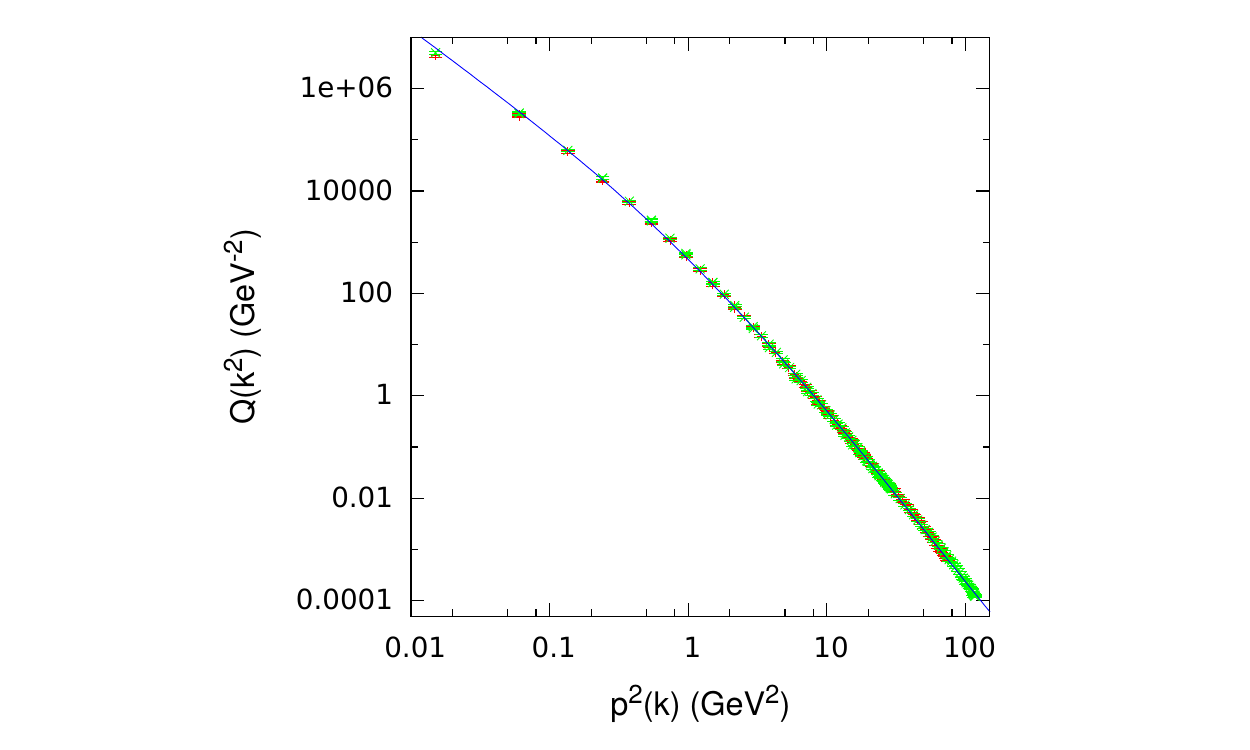}
\vskip 1mm
\includegraphics[trim=55 0 40 0, clip, scale=1.05]{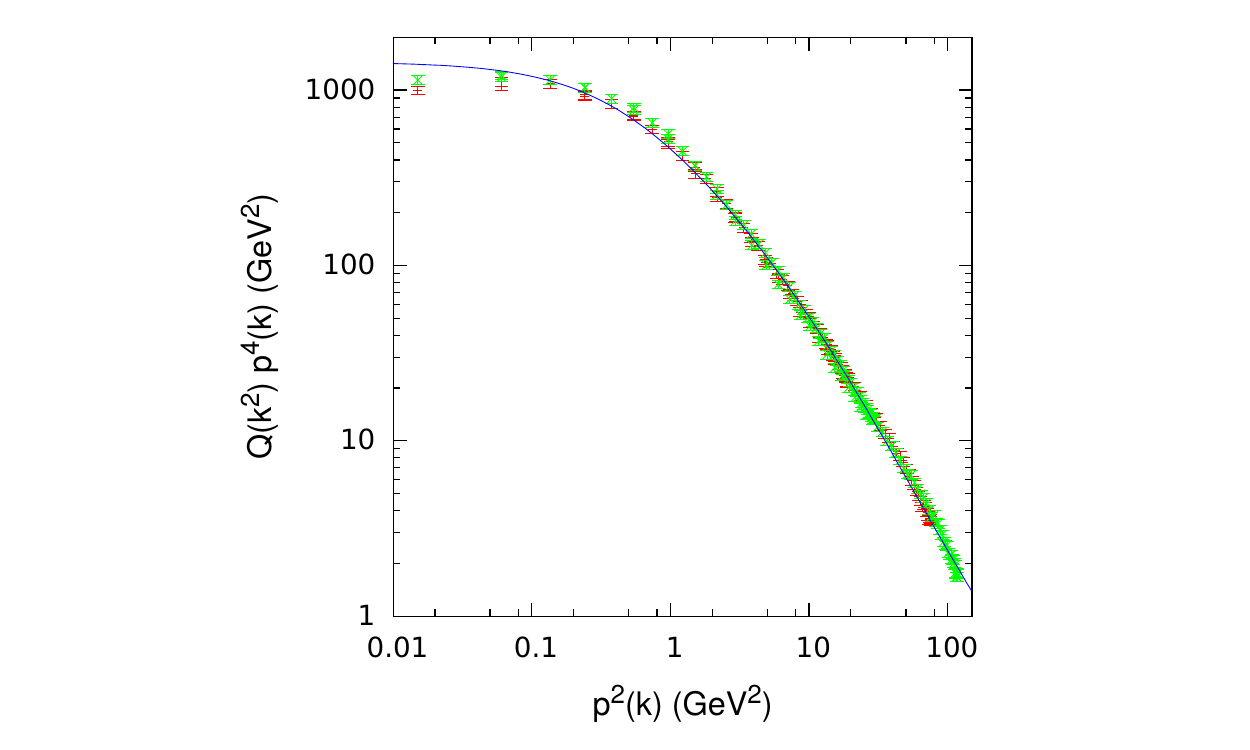}
\caption{\label{fig:fit3} The Bose-ghost propagator $Q(k^2)$ ({\bf top})
[see Eq.\ (\ref{eq:Q})] and the rescaled Bose-ghost propagator
$Q(k^2) p^4(k)$ ({\bf bottom}), as a function of the improved momentum
squared $p^2(k)$ [see Eqs.\ (\ref{eq:pim}) and (\ref{eq:pk})].
We plot data for $\beta_2$, $V = 96^4$ (red, $\times$) and
$\beta_3$, $V = 120^4$ (green, $*$), after applying a matching
procedure \cite{Leinweber:1998uu,Cucchieri:2003di} to the former set of
data. Here we use the sources defined in Eq.\ (\ref{eq:Blatt2}).
We also plot, for $V = 120^4$, a fit using Eq.\ (\ref{eq:fit}) and the
parameters in Table \ref{tab:fits}, with $c = 132(11)$. Note the logarithmic
scale on both axes.
}
\end{figure}


The inversion of the FP matrix ${\cal M}^{ab}(x,y)$ is performed using a
conjugate-gradient method, accelerated by even/odd preconditioning. After
indicating with
\begin{equation}
\widetilde{R}^{a c}_{\mu}(k) \, = \, \frac{1}{\sqrt{V}}
         \sum_x R^{a c}_{\mu}(x) \, e^{2 \pi i k \cdot x / N}
\label{eq:Rk}
\end{equation}
the Fourier transform of the outcome $R^{a c}_{\mu}(x)$ of the numerical
inversion [see Eq.\ (\ref{eq:Rfunc})], it is clear that we can evaluate
the Bose-ghost propagator [see Eq.\ (\ref{eq:Qprop})] in momentum space by
considering
\begin{equation}
Q^{abcd}_{\mu \nu}(k) \, \equiv \, \Re \, \left\{ \,
                  \widetilde{R}^{a b}_{\mu}(k) \,
                   \widetilde{R}^{c d}_{\nu}(-k)
                        \, \right\} \; .
\end{equation}
(In the above equations, $N$ is the lattice side, $k$ is the wave vector
with components $k_{\mu} = 0, 1, \ldots, N-1$ and $\Re$ indicates the
real part of the expression within brackets.) Then, by contracting the
color indices $b, d$ and the Lorentz indices $\mu, \nu$, we can write
[see Eq.\ref{eq:Qabcd})]
\begin{equation}
Q^{a c}(k) \, \equiv \, Q^{abcb}_{\mu \mu}(k) \, \equiv \,
    \delta^{a c} \, P_{\mu \mu}(k) \, Q(k^2) \; ,
\label{eq:Q}
\end{equation}
due to global color invariance. The numerical evaluation of the scalar
function $Q(k^2)$, through lattice Monte Carlo simulations, is the
goal of this work.

\vskip 3mm
The function $\widetilde{R}^{a c}_{\mu}(k)$ defined in Eq.\ (\ref{eq:Rk})
has been evaluated ---for the three different choices for the lattice
sources $B^{bc}_{\mu}(x)$ considered here--- for all possible values of
the color indices $a, c$ and of the Lorentz index $\mu$, and for two
types of momenta, namely wave vectors whose components\footnote{For the
wave vectors $(0,0,0,k)$ we did not consider other possible permutations
of the components.} are $(0,0,0,k)$ and $(k,k,k,k)$, with
$k = 1, 2, \ldots, N/2$. This gives $N$ different values for the
momentum $p$. Note that the null momentum trivially gives
$\widetilde{R}^{a c}_{\mu}(0) = 0$. Indeed, if we indicate with
$\psi^b_{n}(y)$ the eigenvectors of the FP matrix ${\cal M}^{a b}(x,y)$
and with $\lambda_n$ the corresponding eigenvalues,\footnote{Here, the
index $n = 0, 1, 2,\ldots$ denotes the different eigenvalues, possibly
degenerate, of the FP matrix.} then we can write
\begin{equation}
( {\cal M}^{-1} )^{ab}(x,y) \, = \, \sum_{n \neq 0} \frac{\psi^a_{n}(x)
\, \left[\psi^b_{n}(y)\right]^{*}}{\lambda_n} \; ,
\label{eq:M-1}
\end{equation}
where $^{*}$ indicates complex conjugation and $\lambda_0 = 0$ is 
the trivial null eigenvalue, corresponding to constant eigenvectors.
This implies
\begin{eqnarray}
\widetilde{R}^{a c}_{\mu}(k) & = & \frac{1}{\sqrt{V}}
         \sum_{n \neq 0} \frac{1}{\lambda_n} \,
        \left( \left\{ \sum_z \, B^{ec}_{\mu}(z)
                  \, \left[\psi^e_{n}(z)\right]^{*}
         \right\} \right. \nonumber \\[2mm]
  & & \qquad \;\; \left. \left[ \sum_x \, \psi^a_{n}(x)
                \, e^{2 \pi i k \cdot x / N} \right] \,
              \right) \; .
\end{eqnarray}
By recalling that eigenvectors corresponding to distinct eigenvalues 
of symmetric matrices are orthogonal, i.e.\ $\psi_0 = constant$
is orthogonal to the eigenvectors $\psi^a_{n}(x)$ with $n \neq 0$,
we have
\begin{equation}
\sum_x \, \psi^a_{n}(x) \, = \, 0
\end{equation}
for every $n\neq 0$ and therefore $\widetilde{R}^{a c}_{\mu}(k=0) = 0$.

Our numerical code is parallelized using {\tt MPI} and {\tt OpenMP}. We
always use four {\tt OpenMP} threads for each {\tt MPI} task and the
{\tt 4-way Symmetrical Multiprocessing} mode ({\tt SMP}) for the runs on
the {\tt Blue Gene/P} supercomputer at Rice University. The total computing
time was about 2.2 millions of CPU-hours. Further details on the
implementation of the numerical simulations can be found in Ref.\
\cite{Cucchieri:2003zx}.


\section{Numerical Results}

In this section we present the numerical results for the scalar function
$Q(k^2)$, defined in Eq.\ (\ref{eq:Q}) above. In all cases the data points
represent averages over gauge configurations and error bars correspond to
one standard deviation (we consider the statistical error only). Also, in
the plots, all quantities are in physical units. Let us stress that,
compared to the results reported in Refs.\ \cite{Cucchieri:2014via,
Cucchieri:2014xfa}, the data shown here have been divided by an additional
factor 3.

We first investigate the effect of rotational-symmetry breaking on our results,
by plotting the data for $Q(k^2)$ as a function of two different definitions
of the lattice momenta, i.e.\ the usual unimproved definition
\begin{equation}
p^2(k)  \, = \, \sum_{\mu} p_{\mu}^2
\label{eq:pun}
\end{equation}
and the improved definition \cite{Ma:1999kn}
\begin{equation}
p^2(k)  \, = \, \sum_{\mu} \left[ \, p_{\mu}^2(k) + \frac{p_{\mu}^4(k)}{12}
                           \, \right] \; ,
\label{eq:pim}
\end{equation}
where 
\begin{equation}
p_{\mu}(k) \, = \, 2 \sin{\left(\frac{\pi k_{\mu}}{N}\right)} \; .
\label{eq:pk}
\end{equation}
In Figs.\ \ref{fig:rot0} and \ref{fig:rot} we show our data for $Q(k^2)$
---respectively using the lattice definition of the sources $B^{bc}_{\mu}(x)$ 
given in Eqs.\ (\ref{eq:Blatt1}) and (\ref{eq:Blatt2})--- as a function of
the unimproved and of the improved momentum squared $p^2(k)$.
As one can see\footnote{Recall that we expect a factor difference of 4
between the data in these two figures.}, the improved definition
(\ref{eq:pim})--(\ref{eq:pk}) makes the behavior of the propagator
smoother at large momenta, allowing a better fit to the data. We also check
(see Fig.\ \ref{fig:dis}) that our results do not depend on the choice of
the lattice definition for the source 
(see discussion in the previous section).\footnote{From now on 
we will only show data obtained using the trivial discretization 
(\ref{eq:Blatt2}) for the sources $B^{bc}_{\mu}(x)$.}
Finally, in Figs.\ \ref{fig:vol2} and \ref{fig:vol},
which refer respectively to the cases $\beta_0$ and $\beta_1$ at about
the same physical volume, we consider the extrapolation to the 
infinite-volume limit. It is clear that
the use of larger lattice volumes does not modify the behavior of the
propagator, i.e.\ finite-size effects ---at a given lattice momentum $p$---
are essentially negligible. Thus, large volumes are relevant only to clarify
the IR behavior of the Bose-ghost propagator.

\vskip 3mm

In order to extrapolate our data to the continuum limit, we compare data
obtained at different $\beta$ values, using the largest physical volumes
available for comparison.
In particular, in the top plot of Fig.\ \ref{fig:fit} we show the data at
$\beta_0$ with $V = 48^4$ and at $\beta_1$ with $V = 72^4$,
which correspond to the same physical volume,
after rescaling the data at $\beta_0$ using the matching technique
described in Ref.\ \cite{Leinweber:1998uu,Cucchieri:2003di}. Similarly, in
the top plot of Fig.\ \ref{fig:fit2} we compare the data for $\beta_1$
with $V = 72^4$ and $\beta_2$ with $V = 96^4$, and
in the top plot of Fig.\ \ref{fig:fit3} we compare the data for $\beta_2$
with $V = 96^4$ and $\beta_3$ with $V = 120^4$,
always applying a rescaling to the coarser set of data.\footnote{Let us
recall that all these lattice volumes correspond to a physical volume of
about $(10.1 \, fm)^4$.} The data clearly scale quite well, even though
small deviations are observable in the IR limit [see the bottom plots in
Figs.\ \ref{fig:fit}, \ref{fig:fit2} and \ref{fig:fit3}, where we
show the Bose-ghost propagator $Q(k^2)$ multiplied by $p^4(k)$].
We thus observe discretization effects for the coarser lattices.
Nevertheless, these effects decrease as the lattice spacing $a$ goes
to zero. Indeed, the ratios between the finer and the coarser data ---at
the smallest momentum $p^2(k) \approx 0.015 \, GeV^2$--- in the various cases
are, respectively,
equal to 1.66(7), 1.25(9) and 1.14(9) in the plots shown in
Figs.\ \ref{fig:fit}--\ref{fig:fit3}, with errors obtained from
propagation of errors.

As done in Refs.\ \cite{Cucchieri:2014via,Cucchieri:2014xfa}, we also fit
the data using the fitting function\footnote{We note that, in order to
improve the stability of the fit, we impose some parameters to be positive,
by forcing them to be squares.}
\begin{equation}
\label{eq:fit}
f(p^2) \, = \, \frac{c}{p^4} \, \frac{p^2 + s}{p^4 \, + \,
                           u^2 p^2 \, + \, t^2 } \; ,
\end{equation}
which is based on the analysis carried on in Refs.\ \cite{Zwanziger:2009je,
Zwanziger:2010iz}, i.e.\ on the relation (obtained using a cluster
decomposition)
\begin{equation}
Q(p^2) \, \sim \, g_0^2 \, G^2(p^2) \, D(p^2) \; ,
\label{eq:cluster}
\end{equation}
where $D(p^2)$ is the gluon propagator and $G(p^2)$ is the ghost propagator.
Then, one can view the above fitting function as generated by an IR-free FP
ghost propagator $G(p^2) \sim 1/p^2$ and by a massive gluon propagator
$D(p^2)$ \cite{Cucchieri:2007rg,Cucchieri:2008fc,Boucaud:2011ug}. The fit
describes the data quite well (see the $\chi^2/\mbox{d.o.f.}$ values in
Table \ref{tab:fits}), even though in the IR limit there is a small discrepancy
between the data and the fitting function considered (see bottom plots in
Figs.\ \ref{fig:fit}--\ref{fig:fit3}).

Let us note that the
fitted value for the parameter $c$ is somewhat arbitrary, since one can
always fix a renormalization condition at a given scale $p^2 = \mu^2$, which
in turn yields a rescaling of the Bose-ghost propagator by a global factor.
One should also note that, from Eqs.\ (\ref{eq:Qprop}) and (\ref{eq:Rfunc}),
it is clear that the propagator $Q(p^2)$ evaluated in this work has a
renormalization constant $Z_Q$ equal to one \cite{Zwanziger:2009je} in the
so-called Taylor scheme \cite{Boucaud:2008gn} and in the algebraic
renormalization scheme \cite{Vandersickel:2012tz}. This implies that $Z_Q$
is also finite in any renormalization scheme (see e.g.\ Refs.\
\cite{Boucaud:2008gn,Boucaud:2005ce} for a discussion on this issue).


\setlength{\tabcolsep}{2.5pt}
\begin{table}[b]
\begin{tabular}{| c | c | c | c | c | c |}
\hline
$V = N^4$      & $\beta$    & $t \, (GeV^2)$ & $u \, (GeV)$ &
$s \, (GeV^2)$ & $\chi^2 / \mbox{d.o.f.}$ \\
\hline \hline
$\,48^4\,$ & $\beta_0$        & 2.3(0.2) & 1.5(0.2) & 10.9(3.4) &
6.57 \\ \hline
$\,64^4\,$ & $\beta_0$        & 2.2(0.2) & 1.5(0.2) & 8.7(2.4) &
4.06 \\ \hline
$\,72^4\,$ & $\beta_1$ & 3.2(0.3) & 3.6(0.4) & 49(14) &
2.45 \\ \hline
$\,96^4\,$ & $\beta_2$ & 3.0(0.1) & 3.9(0.2) & 57.8(9.5) &
1.12 \\ \hline
$\,120^4\,$ & $\beta_3$ & 3.3(0.2) & 4.8(0.3) & 121(21) &
1.98 \\ \hline
\end{tabular}
\caption{
Parameters $t$, $u$ and $s$ from a fit of $f(p^2)$ in Eq.\ (\ref{eq:fit})
to the data. Errors in parentheses correspond to one standard
deviation.  The number of degrees of freedom (dof) is always $N-4$.
We also show the reduced chi-squared $\chi^2 / \mbox{dof}$. Fits
have been done using {\tt gnuplot}.
\label{tab:fits}}
\end{table}

\setlength{\tabcolsep}{2.5pt}
\begin{table}[b]
\begin{tabular}{| c | c | c | c | c | c |}
\hline
$V = N^4$      & $\beta$    & $v \, (GeV^2)$ & $w \, (GeV^2)$ &
$b$ or $\alpha_-$ & type \\ \hline \hline
$\,48^4\,$ & $\beta_0$         & 1.1(0.3)  & 2.0(0.2)    & 4.8(0.1)
& 1 \\ \hline
$\,64^4\,$ & $\beta_0$         & 1.1(0.3)  & 1.9(0.2)    & 4.0(0.1) &
1 \\ \hline
$\,72^4\,$ & $\beta_1$  & 6.5(1.4)  & 5.6(0.2)    &
4.27(0.03) & -1 \\ \hline
$\,96^4\,$ & $\beta_2$  & 7.6(0.8)  & 6.99(0.04)  &
4.091(0.007)  & -1\\ \hline
$\,120^4\,$ & $\beta_3$ & 11.5(1.4) & 11.04(0.06) &
5.460(0.009) & -1 \\ \hline
\end{tabular}
\caption{
Pole parameters [see Eqs.\ (\ref{eq:fitpoles})--(\ref{eq:coeffs3})]
for the fitting function $f(p^2)$, defined in Eq.\ (\ref{eq:fit}). In
the last column we report the type of poles obtained: the value 1
indicates complex-conjugate poles and the value -1 indicates real
poles. Errors in parentheses have been obtained using a Monte Carlo
error analysis (with 10000 samples).
\label{tab:poles}}
\end{table}

\begin{figure}[t]
\centering
\vskip -2mm
\includegraphics[trim=55 0 40 0, clip, scale=1.05]{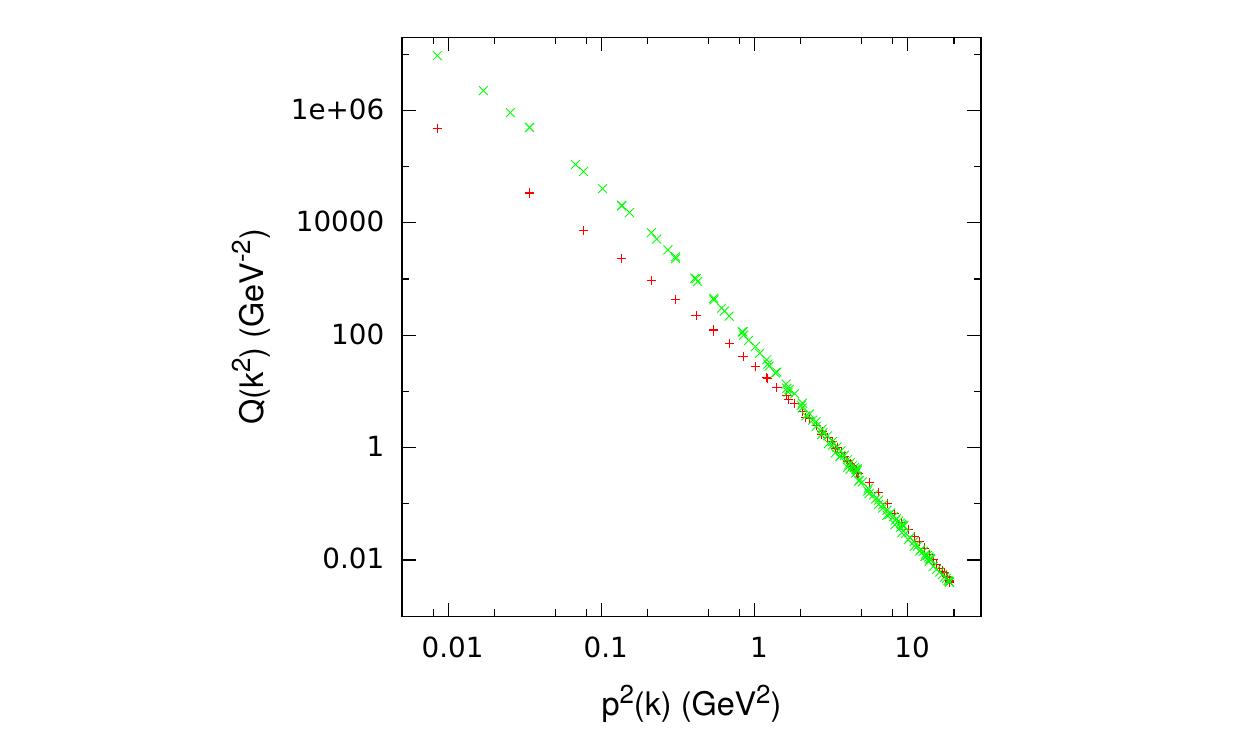}
\caption{\label{fig:compar} The Bose-ghost propagator $Q(k^2)$
(red, $+$) ---defined in Eq.\ (\ref{eq:Q})--- and the product
$g_0^2 \, G^2(p^2) \, D(p^2)$ (green, $\times$) as a function of
the improved momentum squared $p^2(k)$ [see Eqs.\ (\ref{eq:pim})
e (\ref{eq:pk})] for the lattice volume $V = 64^4$ at $\beta_0$.
The data of the Bose-ghost propagator have been
rescaled in order to agree with the data of the product $g_0^2
\, G^2(p^2) \, D(p^2)$ at the largest momentum.
Here we use the sources defined in Eq.\ (\ref{eq:Blatt2}).
Note the logarithmic scale on both axes.
}
\end{figure}

\begin{figure}[t]
\centering
\vskip -2mm
\includegraphics[trim=55 0 40 0, clip, scale=1.05]{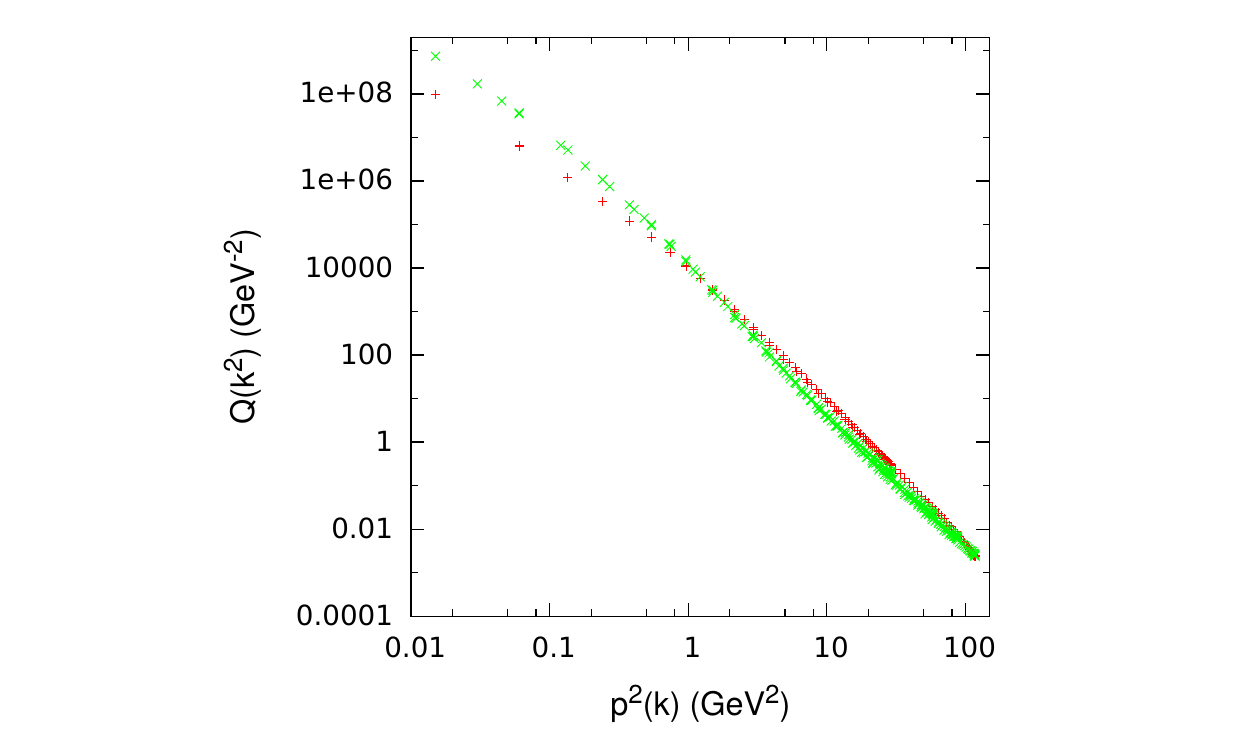}
\caption{\label{fig:compar2} The Bose-ghost propagator $Q(k^2)$
(red, $+$) ---defined in Eq.\ (\ref{eq:Q})--- and the product
$g_0^2 \, G^2(p^2) \, D(p^2)$ (green, $\times$) as a function of
the improved momentum squared $p^2(k)$ [see Eqs.\ (\ref{eq:pim})
e (\ref{eq:pk})] for the lattice volume $V = 120^4$ at $\beta_3$.
The data of the Bose-ghost propagator have been
rescaled in order to agree with the data of the product $g_0^2
\, G^2(p^2) \, D(p^2)$ at the largest momentum.
Here we use the sources defined in Eq.\ (\ref{eq:Blatt2}).
Note the logarithmic scale on both axes.
}
\end{figure}


On the other hand, the parameters $t$, $u$ and $s$ can be related to
the analytic structure of the Bose-ghost propagator $Q(p^2)$. For
example, one could rewrite the fitting function in terms of
a pair of poles, i.e.\
\begin{equation}
\label{eq:fitpoles}
f(p^2) \, = \, \frac{c}{p^4} \,
       \left( \frac{\alpha_+}{p^2+\omega_+^2} \,+\,
              \frac{\alpha_-}{p^2 + \omega_-^2} \right) \; .
\end{equation}
If the poles are complex-conjugate, i.e.\ if $\alpha_{\pm} = 1/2
\pm i b/2 $ and $\omega_{\pm}^2 = v \pm i w $, one has
\begin{equation}
\label{eq:fitpoles2}
f(p^2) \, = \, \frac{c}{p^4} \,
        \frac{p^2 \,+\, v + b w}{
                    p^4 \,+\, 2 v \,p^2 \,+\, v^2 + w^2}
\end{equation}
and
\begin{equation}
s \,=\, v + b w \; , \qquad u^2 \,=\, 2 v \; , \qquad t^2 \,=\,
                 v^2 \, + \, w^2 \; .
\label{eq:coeffs}
\end{equation}
On the contrary, if the poles are real, i.e.\ if $\alpha_{\pm},
\omega_{\pm}^2 = v \pm w \in \mathbb{R}$,
we have
\begin{equation}
\label{eq:fitpoles3}
f(p^2) \, = \, \frac{c}{p^4} \,
       \frac{p^2 \,+\, v +
                \left(\alpha_{-} - \alpha_{+}\right) w}{
                p^4 \,+\, 2 v \,p^2 \,+\, v^2 - w^2}
\end{equation}
and
\begin{eqnarray}
\!\!\!\!\!\!\!
 s &=& v + \left(\alpha_{-} - \alpha_{+} \right) w \; , \qquad \quad \;
              u^2 \,=\, 2 v \; , \\
\!\!\!\!\!\!\!
t^2 &=& v^2 \, - \, w^2 \; , \qquad \qquad
   \alpha_{+} + \alpha_{-} \, = \, 1 \; .
\label{eq:coeffs3}
\end{eqnarray}
Results for these parametrizations for the poles 
are reported in Table \ref{tab:poles}
for the same lattice volumes considered in Table \ref{tab:fits}. (Errors,
shown in parentheses, correspond to one standard deviation and were
obtained using a Monte Carlo error analysis with 10000 samples). We find
that, for the coarsest lattices, i.e.\ at $\beta_0 = 2.2$, these poles are
complex-conjugate, with an imaginary part that is about twice the
corresponding real part. This is in agreement with the results obtained
for the gluon propagator in Ref.\ \cite{Cucchieri:2011ig} at the same
$\beta$ value. On the contrary, for the other three values of $\beta$
considered, we find that these poles are actually real, with $v \approx w$.
In both cases, this fit supports the so-called massive solution of the coupled
Yang-Mills Dyson-Schwinger equations of gluon and ghost propagators (see e.g.\
Refs.\ \cite{Aguilar:2004sw,Aguilar:2006gr,Aguilar:2008xm,Binosi:2009qm,
Aguilar:2011ux,Binosi:2012sj}) and the so-called Refined GZ approach
\cite{Dudal:2008rm,Dudal:2008sp,Dudal:2011gd}. For the cases $V =  48^4$
and $64^4$ at $\beta_0$ we can compare our results in Table \ref{tab:fits}
with the analysis reported in Table II of Ref.\ \cite{Cucchieri:2011ig}
for the gluon propagator. One sees that the fitting parameters for the
Bose-ghost propagator do not seem to relate in a simple way to the
corresponding values obtained by fitting gluon-propagator data. There
is indeed a visible  discrepancy between the Bose-ghost propagator
$Q(p^2)$ and the product $g_0^2 \, G^2(p^2) \, D(p^2)$, as one can
see in Fig.\ \ref{fig:compar} for the lattice volume $V = 64^4$ at
$\beta_0$. This discrepancy seems, however, to decrease at larger
$\beta$ values (see Fig.\ \ref{fig:compar2}).

\begin{figure}[t]
\centering
\vskip -2mm
\includegraphics[trim=55 0 40 0, clip, scale=1.05]{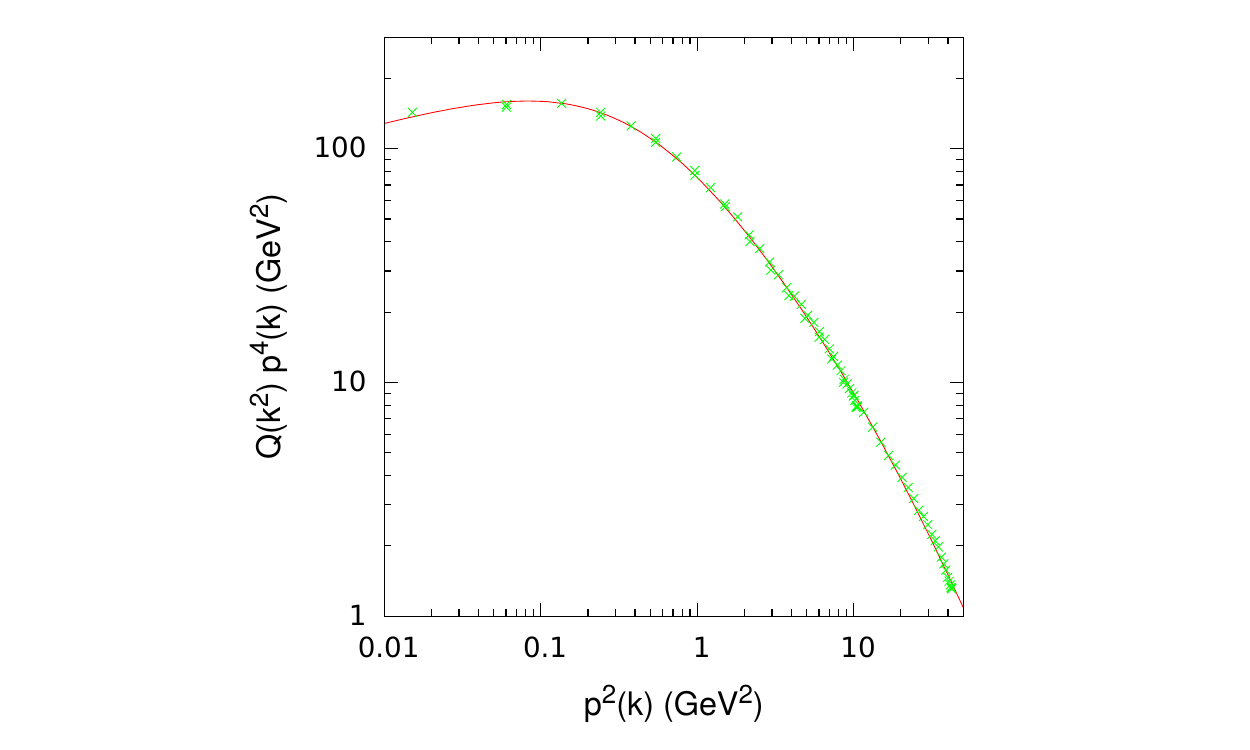}
\caption{\label{fig:fitnew} The rescaled Bose-ghost propagator
$Q(k^2) p^4(k)$ [see Eq.\ (\ref{eq:Q})], as a function of the
improved momentum squared $p^2(k)$ [see Eqs.\ (\ref{eq:pim}) e
(\ref{eq:pk})] for the data at $\beta_1$ and $V = 72^4$.
Here we use the sources defined in Eq.\ (\ref{eq:Blatt2}).
We also plot a fit using Eq.\ (\ref{eq:fit2}) and the
parameters in Table \ref{tab:fits2}, with $c = 15.5(9.1)$.
Note the logarithmic scale on both axes.
}
\end{figure}
\begin{figure}[t]
\centering
\vskip -2mm
\includegraphics[trim=55 0 40 0, clip, scale=1.05]{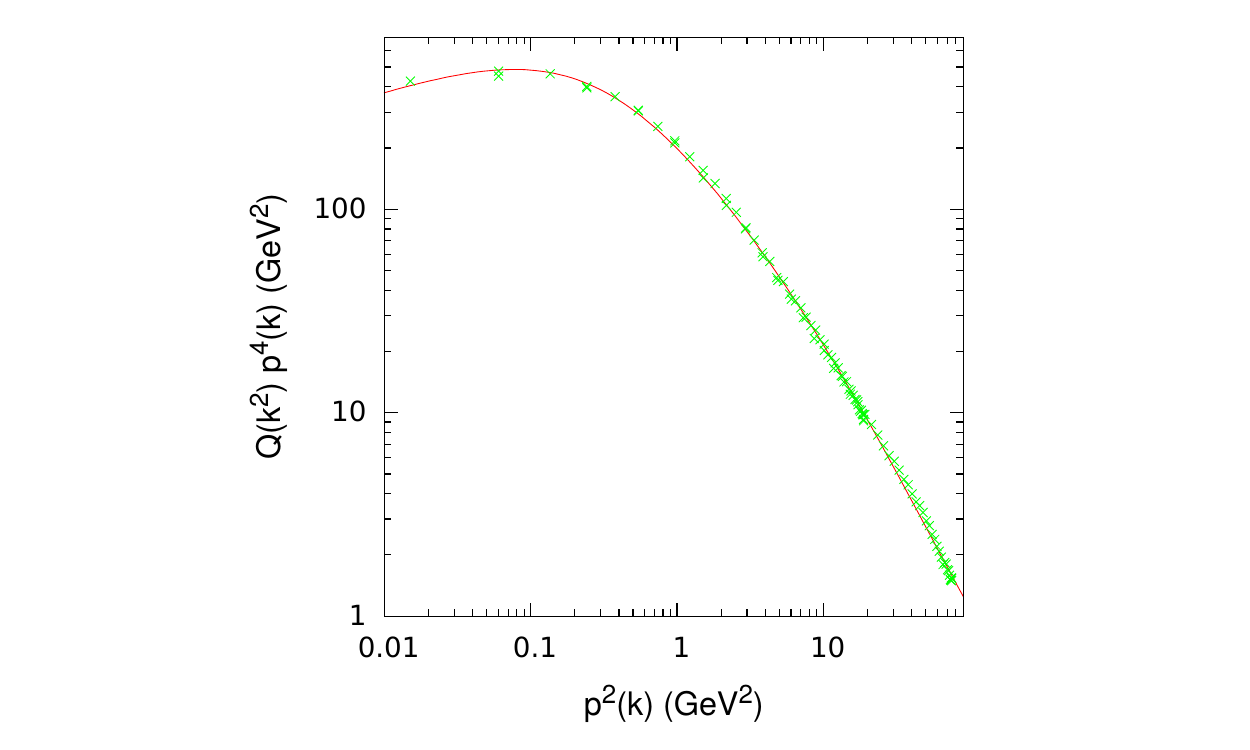}
\caption{\label{fig:fitnew2} The rescaled Bose-ghost propagator
$Q(k^2) p^4(k)$ [see Eq.\ (\ref{eq:Q})], as a function of the
improved momentum squared $p^2(k)$ [see Eqs.\ (\ref{eq:pim}) e
(\ref{eq:pk})] for the data at $\beta_2$ and $V = 96^4$.
Here we use the sources defined in Eq.\ (\ref{eq:Blatt2}).
We also plot a fit using Eq.\ (\ref{eq:fit2}) and the
parameters in Table \ref{tab:fits2}, with $c = 68.4(5.9)$.
Note the logarithmic scale on both axes.
}
\end{figure}
\begin{figure}[t]
\centering
\vskip -2mm
\includegraphics[trim=55 0 40 0, clip, scale=1.05]{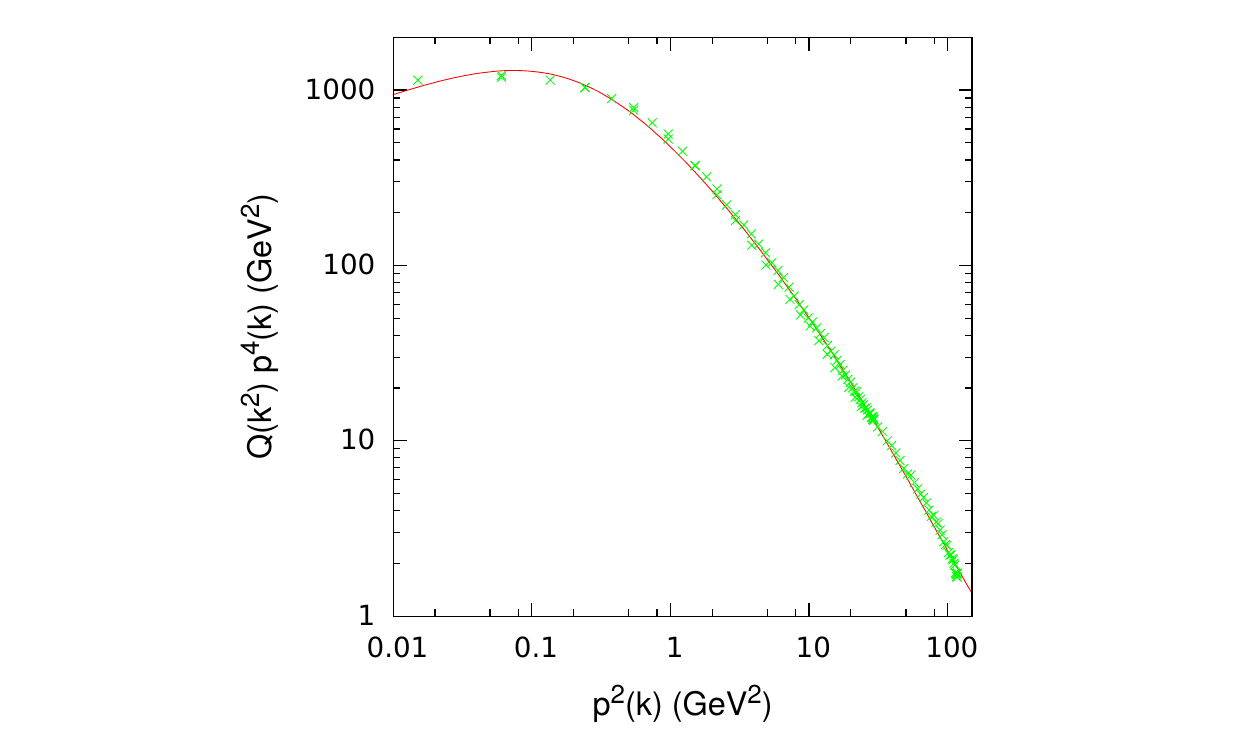}
\caption{\label{fig:fitnew3} The rescaled Bose-ghost propagator
$Q(k^2) p^4(k)$ [see Eq.\ (\ref{eq:Q})], as a function of the
improved momentum squared $p^2(k)$ [see Eqs.\ (\ref{eq:pim}) e
(\ref{eq:pk})] for the data at $\beta_3$ and $V = 120^4$.
Here we use the sources defined in Eq.\ (\ref{eq:Blatt2}).
We also plot a fit using Eq.\ (\ref{eq:fit2}) and the
parameters in Table \ref{tab:fits2}, with $c = 116(12)$.
Note the logarithmic scale on both axes.
}
\end{figure}

Even though the simple Ansatz in Eq.\ (\ref{eq:fit}) above gives a good
description of the data, deviations can be seen in the IR region for
momenta below about $1 \, GeV$, by plotting the quantity $Q(k^2) \,
p^4(k)$ (see bottom plots in Figs.\ \ref{fig:fit}--\ref{fig:fit3}).
We tried to improve our fits, by using more general forms of the
Bose-ghost propagator. In particular, by considering the fitting forms
for the gluon propagator used in Refs.\ \cite{Cucchieri:2003di,
Alkofer:2003jj} we tried to include noninteger exponents in the fitting
function $f(p^2)$. Among the different possibilities considered, the best
results have been obtained with the expression
\begin{equation}
\label{eq:fit2}
f(p^2) \, = \, \frac{c}{p^{4-2\eta}} \, \left( \frac{p^2 + s}{p^4 \, + \,
                           u^2 p^2 \, + \, t^2 } \right)^{1+\eta} \; ,
\end{equation}
which is a natural generalization of Eq.\ (\ref{eq:fit}), while
preserving the ultraviolet behavior $1/p^6$. Also, the above formula
still allows a pole decomposition using Eqs.\ (\ref{eq:coeffs}) and
(\ref{eq:coeffs3}), respectively for the complex-conjugate poles and
for the real poles.\footnote{On the other hand, the new fitting function
makes the identification of the gluon and ghost propagators
in Eq.\ (\ref{eq:cluster}) unclear.} Results for these fits can be seen in
Figs.\ \ref{fig:fitnew}--\ref{fig:fitnew3}, which should be compared
to the corresponding fits in the bottom plots of Figs.\
\ref{fig:fit}--\ref{fig:fit3}. The fitting parameters are reported in
Table \ref{tab:fits2}. As one can see, by comparing Table \ref{tab:fits2}
to Table \ref{tab:fits}, the value of $\chi^2 / \mbox{d.o.f.}$ decreased visibly
with the new fitting form, even though in some cases the fitting parameters
are determined with very large errors (see, in particular, the results for
the parameter $s$). Finally, also in this case we evaluated the
parametrizations of the poles for the fitting curves (see 
Table \ref{tab:poles2}).
We find that, in all cases, the poles are real, with $v \approx w$.


\setlength{\tabcolsep}{2.5pt}
\begin{table}[b]
\begin{tabular}{| c | c | c | c | c | c | c |}
\hline
$V = N^4$      & $\beta$    & $t \, (GeV^2)$ & $u \, (GeV)$ &
$s \, (GeV^2)$ & $\eta$ & $\chi^2 / \mbox{d.o.f.}$ \\
\hline \hline
$\,48^4\,$ & $\beta_0$        & 3.7(0.6) & 4.2(0.8) & 172(299) & 0.19(2) &
2.33 \\ \hline
$\,64^4\,$ & $\beta_0$        & 4.0(0.7) & 4.3(0.9) & 176(342) & 0.16(2) &
1.86 \\ \hline
$\,72^4\,$ & $\beta_1$ & 4.0(0.4) & 6.0(0.8) & 199(143) & 0.19(3) &
1.46 \\ \hline
$\,96^4\,$ & $\beta_2$ & 2.9(0.1) & 5.2(0.3) & 85(15) & 0.24(4) &
0.72 \\ \hline
$\,120^4\,$ & $\beta_3$ & 3.0(0.2) & 6.0(0.4) & 136(23) & 0.30(7) &
1.55 \\ \hline
\end{tabular}
\caption{
Parameters $t$, $u$, $s$ and $\eta$ from a fit of $f(p^2)$ in
(\ref{eq:fit2}) to the data. Errors in parentheses correspond to
one standard deviation. The number of degrees of freedom (dof)
is always $N-5$. We also show the reduced chi-squared $\chi^2 /
\mbox{dof}$. Fits have been done using {\tt gnuplot}.
\label{tab:fits2}}
\end{table}

\setlength{\tabcolsep}{2.5pt}
\begin{table}[b]
\begin{tabular}{| c | c | c | c | c | c |}
\hline
$V = N^4$      & $\beta$    & $v \, (GeV^2)$ & $w \, (GeV^2)$ &
$b$ or $\alpha_+$ & type \\ \hline \hline
$\,48^4\,$ & $\beta_0$         & 8.8(3.4)  & 8.0(0.3)    & 10.69(0.04)
& -1 \\ \hline
$\,64^4\,$ & $\beta_0$         & 9.2(3.9)  & 8.3(0.3)    & 10.50(0.04)
& -1 \\ \hline
$\,72^4\,$ & $\beta_1$  & 18.0(4.8)  & 17.55(0.09)    &
5.66(0.01) & -1 \\ \hline
$\,96^4\,$ & $\beta_2$  & 13.5(1.6)  & 13.21(0.02)  &
3.206(0.004)  & -1\\ \hline
$\,120^4\,$ & $\beta_3$ & 18.0(2.4) & 17.75(0.04) &
3.824(0.006) & -1 \\ \hline
\end{tabular}
\caption{
Pole parameters [see Eqs.\ (\ref{eq:fitpoles})--(\ref{eq:coeffs3})]
for the fitting function $f(p^2)$, defined in Eq.\ (\ref{eq:fit2}). In
the last column we report the type of poles obtained: the value 1
indicates complex-conjugate poles and the value -1 indicates real
poles. Errors in parentheses have been obtained using a Monte Carlo
error analysis (with 10000 samples).
\label{tab:poles2}}
\end{table}


\section{Conclusions}

As explained in the Introduction,
breaking of the BRST symmetry in the GZ approach is linked to a nonzero
value of the
Gribov parameter $\gamma$, which is not assessed directly in lattice
simulations. Nevertheless, this breaking may be also related to a 
nonzero value for the expectation value of a BRST-exact quantity such 
as the Bose-ghost propagator $Q^{abcd}_{\mu \nu}(x,y)$ defined in Eq.\ 
(\ref{eq:defBG}). 
The fact that this quantity can be evaluated on the
lattice ---in much the same way as the ghost propagator 
\cite{Suman:1995zg,Cucchieri:1997dx}--- provides us with
a suitable strategy to study the BRST-symmetry breaking of the GZ action
numerically.
These two manifestations of BRST-symmetry breaking are not independent,
of course, since a nonzero value of $\gamma$ is necessary in both
cases. Indeed, the fact that the lattice gauge-fixing is implemented by a
minimization procedure is already equivalent to a nonzero value of $\gamma$,
while the verification that the Bose-ghost propagator is 
itself nonzero provides nontrivial additional evidence for the breaking.
Note that for $\gamma>1$ the breaking is more pronounced for the
Bose-ghost propagator than for the action, the two being respectively
of order $\gamma^4$ [see e.g.\ Eq.\ (\ref{eq:Qabcd})] and 
$\gamma^2$ [see Eq.\ (\ref{eq:break})].

In this work, we consider for the description of the Bose-ghost
propagator the scalar function $Q(k^2)$
defined in Eq.\ (\ref{eq:Q}), obtained by contracting Lorentz and color
indices in the original propagator.
We recall that this propagator has been proposed as a carrier
of the long-range confining force in minimal Landau gauge
\cite{Zwanziger:2009je,Furui:2009nj,Zwanziger:2010iz}.
We have performed simulations for lattice volumes up to $120^4$ and for
physical lattice extents up to $13.5$ fm, complementing previous 
results reported in \cite{Cucchieri:2014via,Cucchieri:2014xfa}. In
particular, we present a more detailed discussion of the simulations
and we investigate the approach to the infinite-volume and continuum limits.
We find no
significant finite-volume effects in the data. 
As for discretization effects, on the contrary, we observe small 
such effects for the coarser lattices, especially in the IR region.
We also test different
discretizations for the sources $B^{bc}_{\mu}(x)$ used in the inversion
of the FP matrix and find that the data are fairly independent of the
chosen lattice discretization of these sources.

Our results concerning the symmetry breaking and the form of the Bose-ghost
propagator are similar to the previous analysis, i.e.\ we find a $1/p^4$
behavior in the IR regime and a $1/p^6$ behavior at large momenta.
Also, when describing the data by polynomial fits, with the same fitting
forms used in \cite{Cucchieri:2014via,Cucchieri:2014xfa}, we see that
the description is relatively good, improving considerably for the
finer lattices. In particular, plots of the rescaled propagator
show much better agreement with the fit for the finer lattices.
The same does {\em not} hold when using a modified fit with noninteger
exponents as in Eq.\ (\ref{eq:fit2}). Indeed, in this case, although the 
values of $\chi^2/\mbox{d.o.f.}$ are generally better (but the fit has one
extra parameter), the agreement does not get better as one moves to finer
lattices.

Finally, in order to corroborate the results presented here and in Refs.\
\cite{Cucchieri:2014via,Cucchieri:2014xfa}, it would be, of course, important
to evaluate numerically other correlation functions related to the breaking
of the BRST symmetry in the GZ approach and, ultimately, obtain a
lattice estimate of the Gribov parameter $\gamma$.


\section*{Acknowledgments}

The authors thank D.~Dudal and N.~Vandersickel for valuable discussions
and acknowledge partial support from CNPq and USP/COFECUB.
We also would like to acknowledge computing time provided on the Blue
Gene/P supercomputer supported by the Research Computing Support Group
(Rice University) and Laborat\'orio de Computa\c c\~ao Cient\'\i fica
Avan\c cada (Universidade de S\~ao Paulo).


\end{document}